\documentclass{pasj02}
\Received{$\langle$reception date$\rangle$}
\Accepted{$\langle$acception date$\rangle$}
\Published{$\langle$publication date$\rangle$}
\usepackage{array}

\begin{document}

\title{
NinjaSat: Astronomical X-ray CubeSat Observatory
}
%\author{PASJ Editorial Office}%
\author{Toru \textsc{Tamagawa}\altaffilmark{1,2,3}}
\author{Teruaki \textsc{Enoto}\altaffilmark{4,1}}
\author{Takao \textsc{Kitaguchi}\altaffilmark{1}}
\author{Wataru \textsc{Iwakiri}\altaffilmark{5}}
\author{Yo \textsc{Kato}\altaffilmark{1}}
\author{Masaki \textsc{Numazawa}\altaffilmark{6}}
\author{Tatehiro \textsc{Mihara}\altaffilmark{1,2}}
\author{Tomoshi \textsc{Takeda}\altaffilmark{3,1}}
\author{Naoyuki \textsc{Ota}\altaffilmark{3,2}}
\author{Sota \textsc{Watanabe}\altaffilmark{3,1}}
\author{Amira \textsc{Aoyama}\altaffilmark{3,1}}
\author{Satoko \textsc{Iwata}\altaffilmark{3,2}}
\author{Takuya \textsc{Takahashi}\altaffilmark{3,2}}
\author{Kaede \textsc{Yamasaki}\altaffilmark{3,2}}
\author{Chin-Ping \textsc{Hu}\altaffilmark{7,1}}
\author{Hiromitsu \textsc{Takahashi}\altaffilmark{8}}
\author{Yuto \textsc{Yoshida}\altaffilmark{3,2}}
\author{Hiroki \textsc{Sato}\altaffilmark{9,2}}
\author{Shoki \textsc{Hayashi}\altaffilmark{3,2}}
\author{Yuanhui \textsc{Zhou}\altaffilmark{3,2}}
\author{Keisuke \textsc{Uchiyama}\altaffilmark{3,2}}
\author{Arata \textsc{Jujo}\altaffilmark{3,2}}
\author{Hirokazu \textsc{Odaka}\altaffilmark{10}}
\author{Tsubasa \textsc{Tamba}\altaffilmark{11}}
\author{Kentaro \textsc{Taniguchi}\altaffilmark{1}}
\altaffiltext{1}{RIKEN Cluster for Pioneering Research, 2-1 Hirosawa, Wako, Saitama 351-0198, Japan}
\altaffiltext{2}{RIKEN Nishina Center for Accelerator-Based Science, 2-1 Hirosawa, Wako, Saitama 351-0198, Japan}
\altaffiltext{3}{Department of Physics, Tokyo University of Science, 1-3 Kagurazaka, Shinjuku, Tokyo 162-8601, Japan}
\altaffiltext{4}{Department of Physics, Kyoto University, Kitashirakawa Oiwake, Sakyo, Kyoto 606-8502, Japan}
\altaffiltext{5}{International Center for Hadron Astrophysics, Chiba University, 1-33 Yayoi, Inage, Chiba, Chiba 263-8522, Japan}
\altaffiltext{6}{Department of Physics, Tokyo Metropolitan University, 1-1 Minamiosawa, Hachioji, Tokyo 192-0397, Japan}
\altaffiltext{7}{Department of Physics, National Changhua University of Education, Changhua, Changhua 50007, Taiwan}
\altaffiltext{8}{Department of Physics, Hiroshima University, 1-3-1 Kagamiyama, Higashi-Hiroshima, Hiroshima 739-8526, Japan}
\altaffiltext{9}{Department of System Engineering and Science, Shibaura Institute of Technology, 307 Fukasaku, Minuma, Saitama, Saitama 337-8570, Japan}
\altaffiltext{10}{Department of Earth and Space Science, Osaka University, 1-1 Machikaneyama, Toyonaka, Osaka 560-0043, Japan}
\altaffiltext{11}{Institute of Space and Astronautical Science, JAXA, 3-1-1 Yoshinodai, Chuo, Sagamihara, Kanagawa 252-5210, Japan}

\email{tamagawa@riken.jp}

\KeyWords{space vehicles --- space vehicles: instruments --- instrumentation: detectors --- X-rays: general}

\maketitle

\begin{abstract}
NinjaSat is an X-ray CubeSat designed for agile, long-term continuous observations of bright X-ray sources, with the size of 6U ($100\times200\times300$~mm$^3$) and a mass of 8~kg.
NinjaSat is capable of pointing at X-ray sources with an accuracy of less than $0^{\circ}\hspace{-1.0mm}.1$ (2$\sigma$ confidence level) with 3-axis attitude control. 
The satellite bus is a commercially available NanoAvionics M6P, equipped with two non-imaging gas X-ray detectors covering an energy range of 2--50~keV.
A total effective area of 32~cm$^2$ at 6~keV is capable of observing X-ray sources with a flux of approximately 10$^{-10}$~erg~cm$^{-2}$~s$^{-1}$.
The arrival time of each photon can be tagged with a time resolution of 61~$\mu$s.
The two radiation belt monitors continuously measure the fluxes of protons above 5~MeV and electrons above 200~keV trapped in the geomagnetic field, alerting the X-ray detectors when the flux exceeds a threshold. 
The NinjaSat project started in 2020.
Fabrication of the scientific payloads was completed in August 2022, and satellite integration and tests were completed in 2023 July.
NinjaSat was launched into a Sun-synchronous polar orbit at an altitude of about 530~km on 2023 November 11 by the SpaceX Transporter-9 mission. 
After about three months of satellite commissioning and payload verification, we observed the Crab Nebula on 2024 February 9 and successfully detected the 33.8262~ms pulsation from the neutron star. 
With this observation, NinjaSat met the minimum success criterion and stepped forward to scientific observations as initially planned.
By the end of 2024 November, we successfully observed 21 X-ray sources using NinjaSat.
This achievement demonstrates that, with careful target selection, we can conduct scientific observations effectively using CubeSats, contributing to time-domain astronomy.
\end{abstract}

%\pagewiselinenumbers

\section{Introduction}

Over the past 60 years, the field of space science has been primarily driven by national space agencies, such as the National Aeronautics and Space Administration (NASA), the Japan Aerospace Exploration Agency (JAXA), etc., contributing to our understanding of the universe through various discoveries.
As our knowledge expands, the demand for larger and more sensitive astronomical satellites continues to grow.
Consequently, costs are increasing, and production periods are becoming longer.
This trend is unfavorable for maintaining the space science experiment community and for making more discoveries quickly. 
For instance, the opportunities to launch newly developed experimental devices into space are decreasing, and the barriers to entry for newcomers are becoming higher.
In the past decade, however, space utilization and exploration by the private sector have rapidly expanded as system components for fabricating satellites have become smaller and cheaper. 
Low Earth Orbit, in particular, has already been established as a commercial field, with private companies increasingly utilizing it for Earth observations and communications. 
We thought to use these private companies for cost-effective scientific observations in X-ray astronomy and started the small satellite mission NinjaSat \citep{enoto2020, Tamagawa2023SSC}.

The CubeSat standard, most commonly utilized for small or nanosatellites, defines satellite size as a multiple of 1U ($100 \times100\times100$~mm$^3$).
Since the first successful launch of CubeSats in 2003 by several institutions, including XI-IV built by the University of Tokyo~\citep{Funase2019SSC} and CUTE-I built by Tokyo Institute of Technology~\citep{CUTE-I}, the cost performance of small satellites has been widely recognized, leading to a significant increase in the number of CubeSat missions.
Recently, over 100 CubeSats have been launched annually worldwide.
In the early days, the main applications were student education and component testing, but since the late 2000s, commercial use of CubeSat has been expanding.

Several attempts have been made to use CubeSats for space science.
The Miniature X-ray Solar Spectrometer (MinXSS) performed solar spectroscopy in the soft X-ray band~\citep{MinXSS2016}.
In the field of X-ray and gamma-ray astronomy, pioneering missions such as HaloSat~\citep{Kaaret_2019}, which observes soft X-rays from the halo of our Galaxy; PolarLight~\citep{feng2019}, which observes X-ray polarization; and GRBAlpha~\citep{GRBAlpha}, which detects gamma-rays from gamma-ray bursts, have produced significant scientific results.
However, achieving both high pointing accuracy and a sufficiently large effective area is essential to observe dozens to hundreds of X-ray astronomical sources with a small satellite.
To meet these stringent requirements, we have designed and developed the X-ray CubeSat NinjaSat to optimize performance for pointing observations of X-ray sources.
This paper provides a comprehensive overview of the NinjaSat project, detailing the project objectives, satellite system, payload specifications, initial operations, and initial on-orbit calibration results.

\section{Science objectives}

\begin{figure*}
 \begin{center}
  \includegraphics[width=160mm]{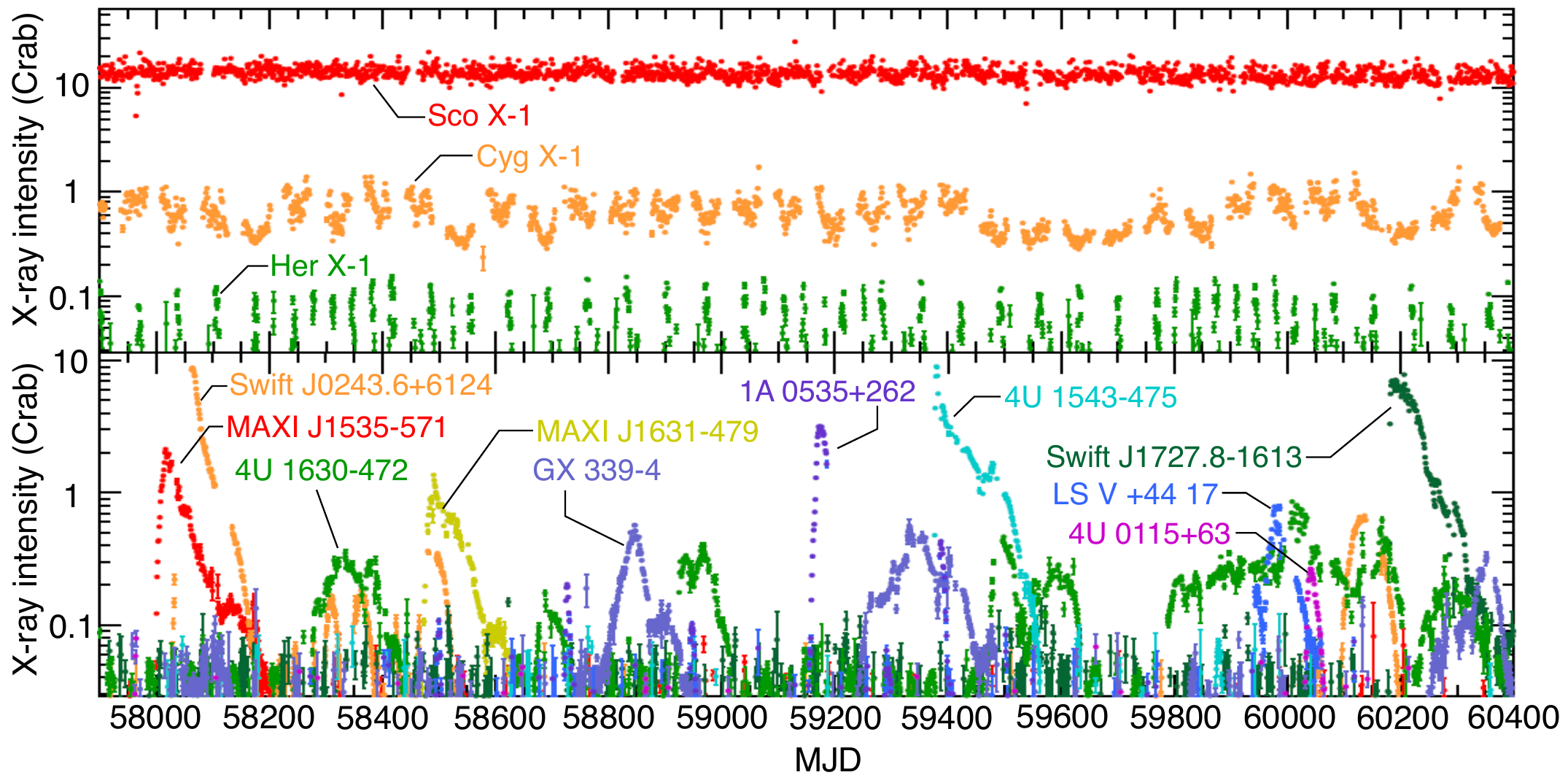}
 \end{center}
 \caption{
The 2--20~keV light curves of bright X-ray sources monitored by MAXI.
 (Upper) Some persistently bright sources: Scorpius X-1, Hercules X-1 (neutron star binaries), and Cygnus X-1 (blackhole). 
 (Lower) The bright transient X-ray sources.
 {Alt text: A stacked graph.}
 }
 \label{fig:maxi_lightcurve}
\end{figure*}

X-ray observations using CubeSats are limited in the resources that can be allocated to scientific instruments, resulting in few missions dedicated to pointing at specific celestial objects.
Unlike optical or infrared bands, X-rays possess high penetrating power, requiring a larger detection volume for effective measurement.
However, even with the size constraints of CubeSats, it is feasible to study the temporal variability of persistently bright or intense transient X-ray sources using smaller detectors, thereby contributing to time-domain astronomy.
Figure~\ref{fig:maxi_lightcurve} shows the light curves of several bright X-ray sources detected by the Gas Slit Camera (GSC; \cite{mihara2011maxigsc}) onboard the all-sky X-ray monitor MAXI \citep{matsuoka2009} attached on the International Space Station over the past six and half years.
Approximately two bright new sources are expected to be observed annually. 
MAXI is capable of observing the entire X-ray sky; however, even when a new X-ray source is discovered, monitoring is limited to about one minute during each 90-minute orbit around the Earth.
NinjaSat has the capability to bridge such observational gaps.
Additionally, there are over 20 persistently bright X-ray sources with an intensity of 0.1~Crab\footnote{One Crab unit is approximately 2.4$\times$10$^{-8}$~erg~cm$^{-2}$~s$^{-1}$ in the 2--10~keV band.} or higher, making them suitable targets for NinjaSat.

As an example of scientific observations using small satellites, we focus on the case of Scorpius (Sco) X-1 \citep{enoto2020}, the brightest X-ray source in the sky, with an intensity of approximately 14~Crab.
Sco X-1 is a low-mass X-ray binary system containing a weakly magnetized neutron star, believed to have been spun up to near the centrifugal breakup frequency in the kHz range through the transfer of angular momentum via mass accretion.
When the neutron star's shape is slightly distorted from the perfect sphere, the fast-spinning neutron star is a potential source of the coherent gravitational wave \citep{1984ApJ...278..345W}. 
The X-ray detection of the kHz twin QPO would provide information on the spin frequency of the neutron star, assuming the beat-frequency model \citep{1996ApJ...469L...1V}. 
Long-term monitoring of this frequency signal could help the LIGO, Virgo, and KAGRA search for coherent gravitational waves from the neutron star to reduce the computational resource \citep{2017PhRvD..95l2003A,2021ApJ...906L..14Z}.
In more general, looking at the scientific advantages of small satellite observations, bright X-ray sources are usually compact objects with time-variability, which can be traced by simultaneous observations at different multi-messenger observations, such as gravitational waves, optical, and radio bands.
%}

The NinjaSat project set the following success criteria.
\begin{itemize}
\item Minimum Success:
Conduct a pointing observation of an X-ray source and detect X-rays from the source.
\item Full Success:
Observe at least two X-ray sources and publish two scientific papers.
\item Extra Success:
Achieve one of the following two. 
1) Conduct simultaneous observations with observatories in other wavelengths and discover something new. 
2) Determine the rotation period of a nearby accreting neutron star which helps to find a coherent gravitational wave from the source.
\end{itemize}
Considering that scientific observations with CubeSats have rarely been conducted so far and are high-risk, the minimum success criterion is set to demonstrate that pointing observations of X-ray sources are possible and that scientific observations with CubeSats are feasible. 
The full success criterion is defined as the achievement of pointing observations and the publication of two scientific papers.
If this criterion is met, we can demonstrate that even a CubeSat-type observatory can contribute to scientific progress.
The extra success criterion is set to demonstrate that even small satellites can achieve discoveries beyond their observational capabilities by collaborating with other observatories.

\section{Spacecraft}

While small satellites are cost-effective, they undeniably face a high risk of failure in space.
The core focus of our work is to observe astronomical X-ray sources using an X-ray detector we have developed rather than building the satellite itself.
To minimize the risk of satellite failure, we decided to outsource the satellite bus fabrication to a company with proven experience in developing and operating CubeSats in space. 
In the NinjaSat project, NanoAvionics, a Lithuanian satellite manufacturer, was selected as the contractor for the satellite bus fabrication, testing, and operation in space, with Mitsui Bussan Aerospace as an intermediary. 
The scientific payloads were fabricated at RIKEN and assembled into the satellite in Lithuania. 
The project, which started in 2020, was delayed about one year due to parts shortages caused by COVID-19, but otherwise, the project proceeded as initially planned.

\begin{figure}
 \begin{center}
  \includegraphics[width=70mm]{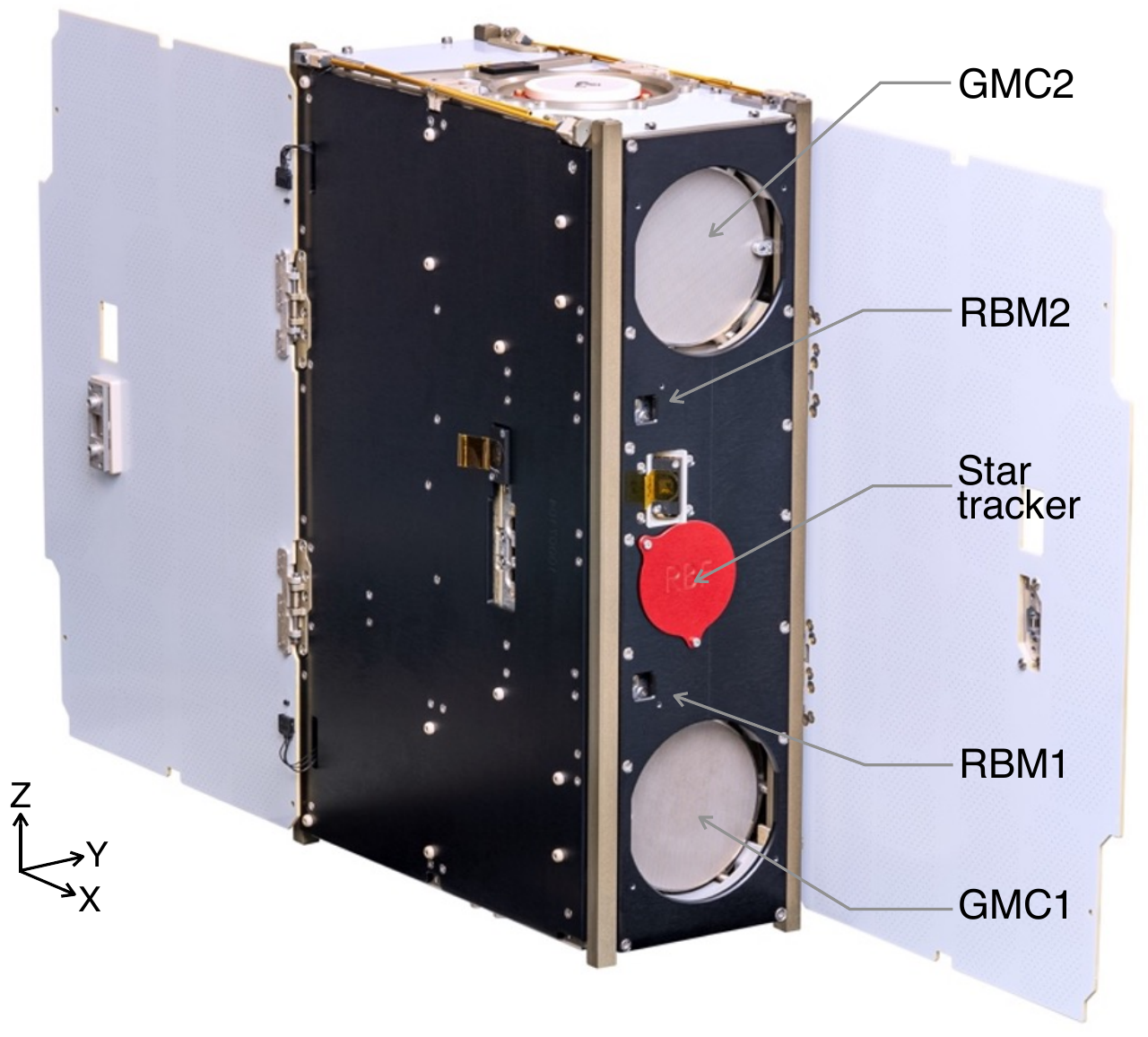}
 \end{center}
 \caption{
 A photograph of NinjaSat with its solar panels deployed.
 The size is the 6U CubeSat standard ($100\times200\times300$~mm$^3$) when the solar panels are stowed.
 The boresights of all payloads (two GMCs and two RBMs) and the star tracker face the same direction (+X).
  {Alt text: A single photograph.}
 }
 \label{fig:nsat_overview}
\end{figure}

NinjaSat is a 6U-size CubeSat ($100\times200\times300$~mm$^2$), as shown in figure~\ref{fig:nsat_overview}, based on the NanoAvionics M6P satellite bus and customized to accommodate our scientific payloads: two Gas Multiplier Counters (GMCs) and two Radiation Belt Monitors (RBMs). 
The boresight of the star tracker is aligned with those of the GMCs and RBMs.
The solar panels were stowed to fit into the 6U size during the launch but were deployed in space. 
The solar battery cells are tiled on the back side of the panel adjacent to the GMCs, preventing sunlight from reaching the GMCs, RBMs, and the star tracker during normal observations.

\begin{figure*}
 \begin{center}
  \includegraphics[width=150mm]{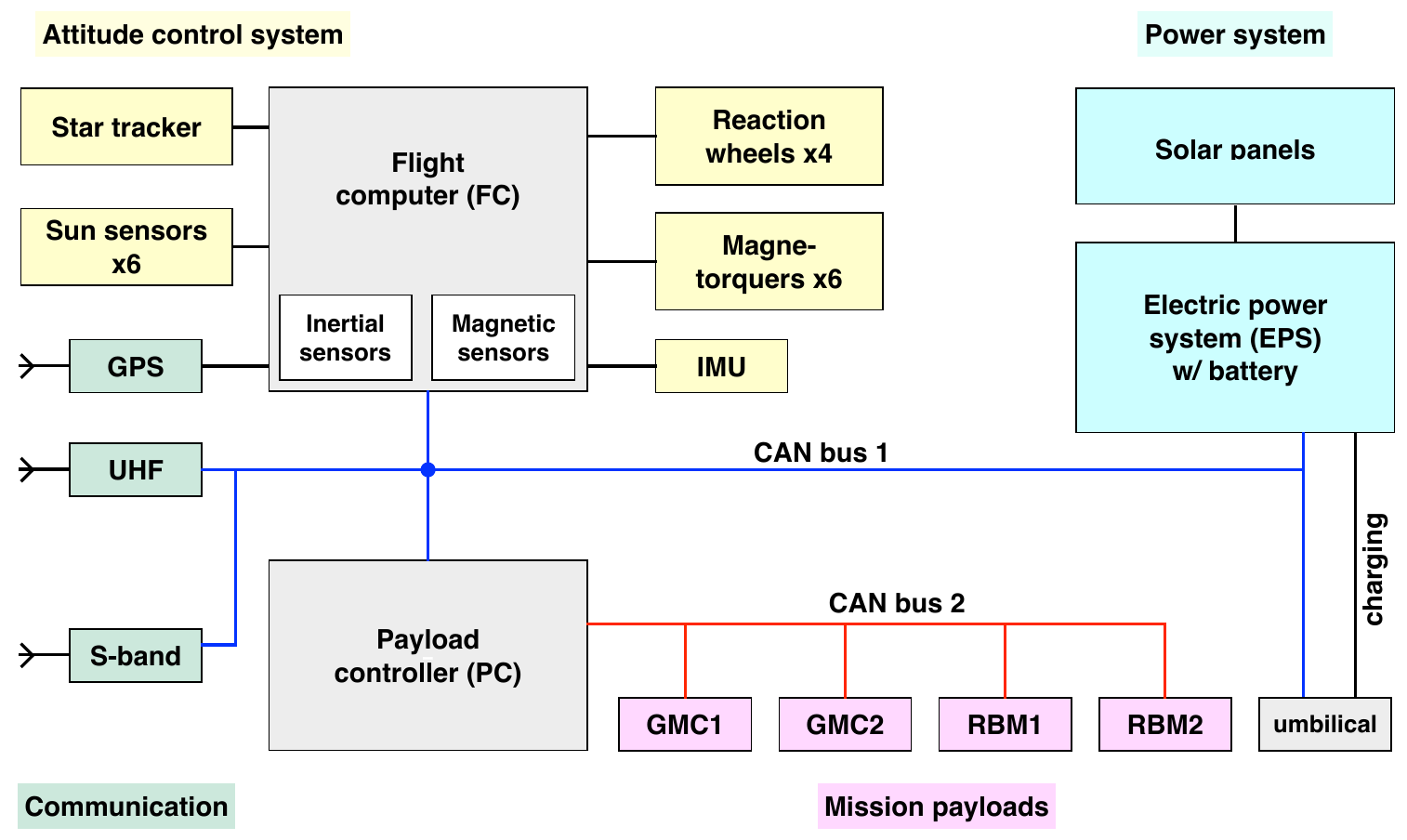}
 \end{center}
 \caption{A functional block diagram of NinjaSat.
 {Alt text: A diagram.}
 }
 \label{fig:nsat_function_diagram}
\end{figure*}

Figure~\ref{fig:nsat_function_diagram} shows the functional block diagram of the satellite.
The Electrical Power System (EPS) manages the power supply, which is essential for satellite operations.
The EPS has the capability to independently control the power supply to all subsystems, turning them ON or OFF as needed. 
Each EPS power supply line is equipped with both hardware and software current limiters; the value of the software current limiter is adjustable via commands.
The scientific payload GMCs and RBMs receive commands from and transmit data to the Payload Controller (PC). 
Since the payloads do not have memory to store data, they send the obtained data to the PC immediately, where the data is stored in 4~GByte serial NAND memory. 
In addition, two 32~GByte SD cards can also be used to store data.
The standard Controller Area Network (CAN) bus utilized by NanoAvionics has been adopted to communicate between satellite subsystems and scientific payloads. 
The internal communication bus of the satellite system (CAN bus 1) is separated from the payload communication bus (CAN bus 2).

NinjaSat performs precision and inertial pointing observations of celestial objects. 
An Attitude Determination and Control System (ADCS) is essential for achieving the required accuracy in these observations. 
To measure the intensity of X-ray sources with an accuracy of 5\%, the ADCS must maintain a pointing accuracy and stability of $0^{\circ}\hspace{-1mm}.1$ relative to the target coordinates.
The ADCS of NinjaSat integrates data from six Sun sensors, a star tracker, an inertial measurement unit (IMU), inertial sensors, magnetic sensors, and date-time information provided by the global positioning system (GPS). 
These sensor inputs are processed through an extended Kalman filter in the flight computer (FC), and the satellite attitude is controlled with four reaction wheels (one of four is redundant) and six magnetorquers (2 per axis).
The ADCS operates in multiple modes, including a high-accuracy determination mode utilizing star tracker data and a coarse determination mode that does not. 
Only the data with the high-accuracy determination mode is used for X-ray source observations.

\begin{table}
  \tbl{Key information of NinjaSat.}{%
  \begin{tabular}{lr}
      \hline
      Parameter & Value or note\\
      \hline
      Size & 6U CubeSat\\
      Satellite bus & NanoAvionics M6P \\
      Power consumption & 16.4~W\footnotemark[$*$] \\
      Weight & 8.14~kg \\
      Pointing accuracy & < $0^{\circ}\hspace{-1mm}.1$ (2$\sigma$ CL)\\
      Maneuver slew speed & $\sim$5$^{\circ}$~s$^{-1}$\\
      Scientific payloads & Two GMCs \& two RBMs\\
      Communication & S-band \& UHF\\
      Ground station & Svalbard (primary) \\
                     & \& Awaruwa (backup) \footnotemark[$\dag$]\\
      Downlink & 60~MB d$^{-1}$ (minimum) \\
      Launch date/time& 2023 Nov 11 10:49 a.m. PST\footnotemark[$\ddag$] \\
      Launch site & Vandenberg Space Force Base\\
      Launcher & SpaceX Falcon 9 block 5\\
      Launch mission & Transporter-9\\
            Orbit & Sun Synchronous Orbit\\ 
      LTDN\footnotemark[$\S$] & 10:32 a.m.\\
      Altitude & 519~km (SMA\footnotemark[$\|$])\\
      Mission lifetime & 2~yr \footnotemark[$\sharp$]\\
      \hline    \end{tabular}}\label{tab:ninjasat_spec}
\begin{tabnote}
\footnotemark[$*$] Daily average.\\ 
\footnotemark[$\dag$] Awaruwa is used only for an emergency case.\\  
\footnotemark[$\ddag$] Pacific Standard Time.  \\ 
\footnotemark[$\S$] Local Time at Descending Node. \\
\footnotemark[$\|$] Semi-major axis altitude. \\
\footnotemark[$\sharp$] 1-year design life plus 1-year extension.\\ 
%\footnotemark[$\dag\dag$]  ... \\ 
\end{tabnote}
\end{table}

The NinjaSat establishes communication with the ground station utilizing Ultra High Frequency (UHF) and S-band frequency.
The primary S-band ground station is Svalbard in the Arctic operated by Kongsberg Satellite Services (KSAT), with the Awarua station in New Zealand serving as a backup.
UHF communication is used for initial satellite deployment and emergency operations, while S-band communication is used for primary uplink and downlink.
Absolute time is obtained from GPS.
The internal Real Time Clock (RTC) of the satellite is synchronized to absolute time according to the 1~Hz Pulse Per Second (PPS) signal provided through the GPS receiver.
The PPS signal is also provided to two GMCs for absolute time correction. 
The umbilical lines include power and CAN bus 1 signal lines, which connect to external Ground Support Electronics (GSE) during ground tests. 
These connections were disabled when the satellite was stowed in the deployer before the launch.
The key information of NinjaSat is summarized in table~\ref{tab:ninjasat_spec}.

\section{Scientific payloads}

\subsection{Gas multiplier counter (GMC)}
% 武田君チェックずみ (2024/11/27)

\begin{figure}
 \begin{center}
  \includegraphics[width=80mm]{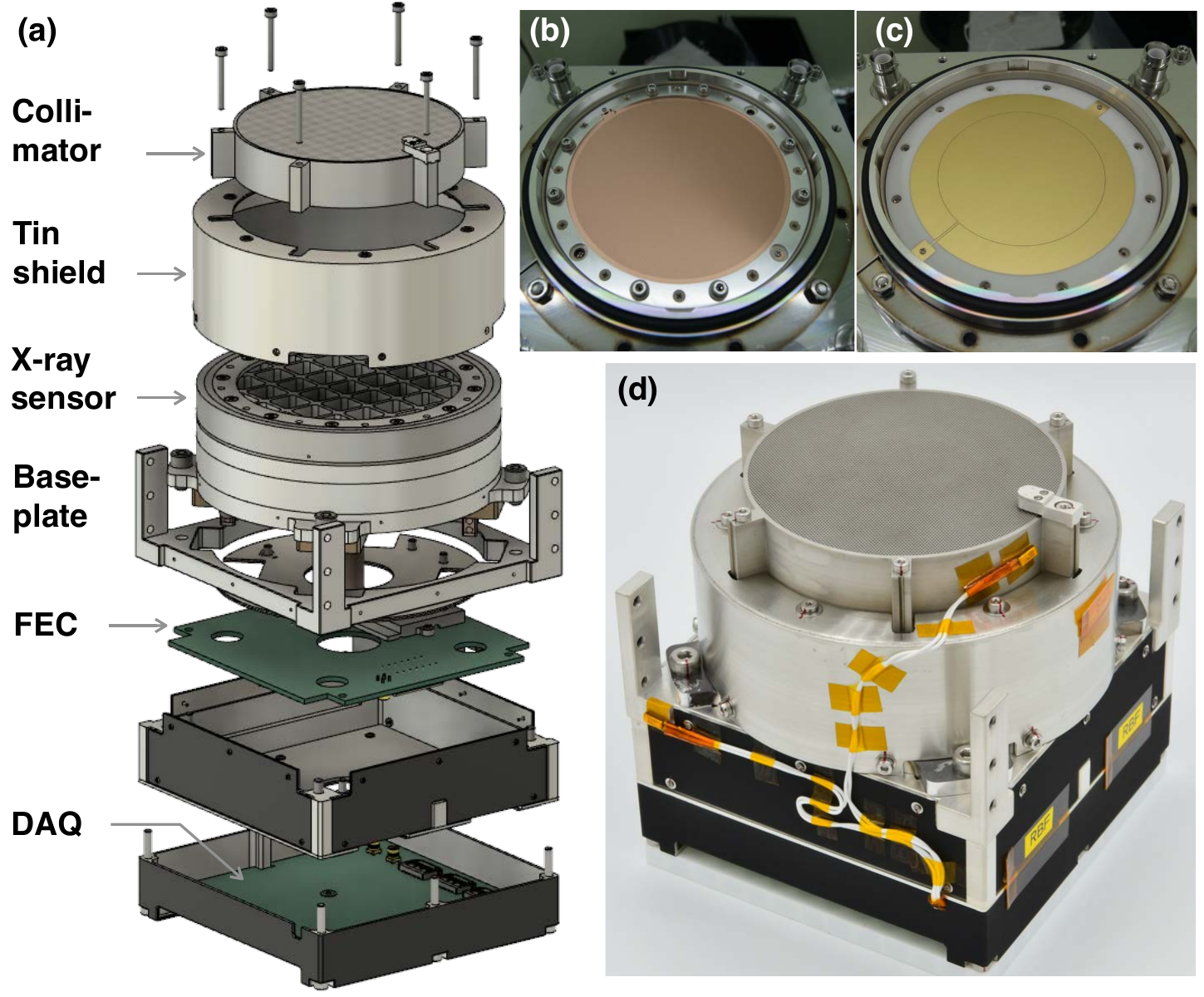}
 \end{center}
 \caption{
(a) An exploded view of the GMC.
(b) A photograph of the Gas Electron Multiplier.
(c) A photograph of readout electrodes.
(d) A photograph of a flight model GMC.
The size of the GMC is fit to 1U (100$\times$100$\times$100~mm$^3$).
{Alt text: A graph consists of one illustration and three photographs.}
}
 \label{fig:nsat_gmc}
\end{figure}

\begin{figure*}
 \begin{center}
  \includegraphics[width=160mm]{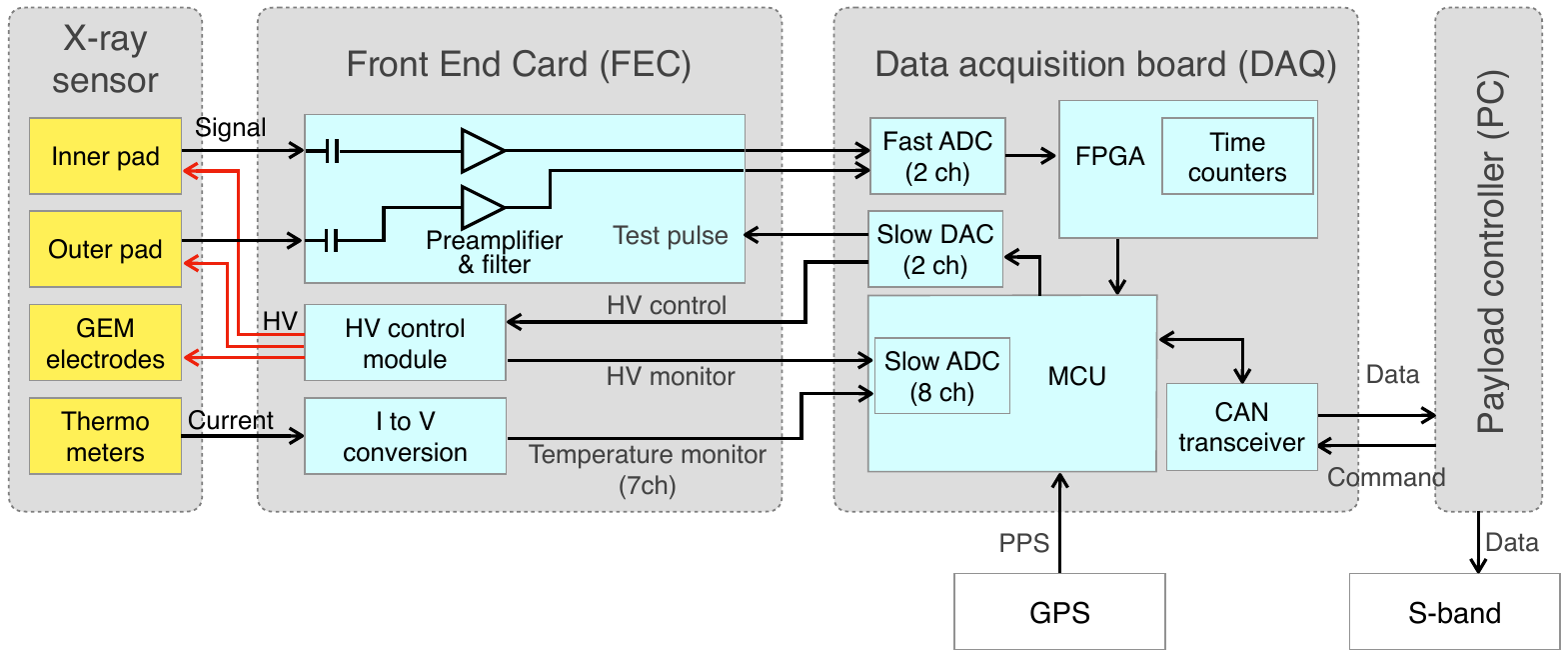}
 \end{center}
 \caption{A functional block diagram of the GMC.
 {Alt text: A diagram.}
 }
 \label{fig:gmc_block_diagram}
\end{figure*}

The GMC is a non-imaging X-ray detector \citep{Takeda2023SSC} and serves as the only astronomical observation instrument onboard NinjaSat. 
An exploded view of the GMC is shown in figure~\ref{fig:nsat_gmc}a.
The GMC consists of a collimator designed to narrow the field of view (FoV), a xenon (Xe)-based gas X-ray proportional counter, an analog signal processing and a high-voltage supply board (Front End Card; FEC), and a digital signal processing board (DAQ).
An attachment to the satellite frame is a base plate.
A tin (Sn) shield is employed to prevent high-energy cosmic diffuse X-rays from outside the FoV.
The FEC and DAQ boards are enclosed within an aluminum shield box to prevent electromagnetic noise.
The functional block diagram of the GMC is shown in figure~\ref{fig:gmc_block_diagram}. 
The parameters characterizing the GMC are presented in table~\ref{tab:gmc_fact}.

\begin{table}
  \tbl{GMC parameters and specifications.}{%
  \begin{tabular}{lr}
      \hline
      Parameter & Value or note\\ 
      \hline
      Size & 1U (100$\times$100$\times$100~mm$^3$)\\
      Weight & 1.2~kg\\
      Operational temperature & $-10^{\circ}$C to $+30^{\circ}$C\\
      Power consumption & 2.0~W (maximum)\\
      Voltage & +5~V (regulated)\\
      Applied high voltages & \\
      \hspace{5mm} Drift plane (Be window) & 0~V \\
      \hspace{5mm} GEM cathode & 602 / 597~V\footnotemark[$*$]\\
      \hspace{5mm} GEM anode   & 1192 / 1194~V\footnotemark[$*$]\\
      \hspace{5mm} Readout pads & 1790 / 1791~V\footnotemark[$*$]\\
      X-ray window & 100~$\mu$m-thick Be\\
      Spacing &\\
      \hspace{5mm} Be window to GEM cathode & 15.45~mm\\
      \hspace{5mm} GEM anode to readout pads & 1.375~mm\\
      GEM & \\
      \hspace{5mm} Diameter & 70~mm\\
      \hspace{5mm} Insulator thickness &  100~$\mu$m\\      
      \hspace{5mm} Hole pitch & 140~$\mu$m\\
      \hspace{5mm} Hole diameter & 70~$\mu$m\\
      Radout pads & \\
      \hspace{5mm} Inner pad diameter & 50~mm\\
      \hspace{5mm} Outer pad diameter & 67~mm\\
      Gas &\\
      \hspace{5mm} Mixture & Xe (75\%), Ar (24\%), DME (1\%) \footnotemark[$\dag$]\\
      \hspace{5mm} Pressure & 1.2~atm at 0$^{\circ}$C\\
      Energy range & 2--50~keV\\
      Effective area & 16~cm$^2$ at 6~keV\\
      CXB shield & 50~$\mu$m-thick Mo and\\
                 & 800 or 500~$\mu$m-thick Sn\footnotemark[$\ddag$]\\
      Collimator & \\
      \hspace{5mm} hole shape & hexagonal\\
      \hspace{5mm} opening fraction & 0.69\\
      \hspace{5mm} FoV & $2^{\circ}\hspace{-1.0mm}.1$ (FWHM)\\
      \hline
    \end{tabular}}\label{tab:gmc_fact}
\begin{tabnote}
\footnotemark[$*$] Applied voltages to GMC1 / GMC2.\\ 
\footnotemark[$\dag$] The values represent volume percentages.\\
\footnotemark[$\ddag$] Details of the Sn shields are described in the text. \\
%\footnotemark[$\ddag$]  ... \\ 
%\footnotemark[$\S$]  ... \\ 
%\footnotemark[$\|$]  ... \\
%\footnotemark[$\sharp$]  ... \\  
%\footnotemark[$**$]  ... \\ 
%\footnotemark[$\dag\dag$]  ... \\ 
\end{tabnote}
\end{table}

\subsubsection{Collimator}

\begin{figure}
 \begin{center}
  \includegraphics[width=80mm]{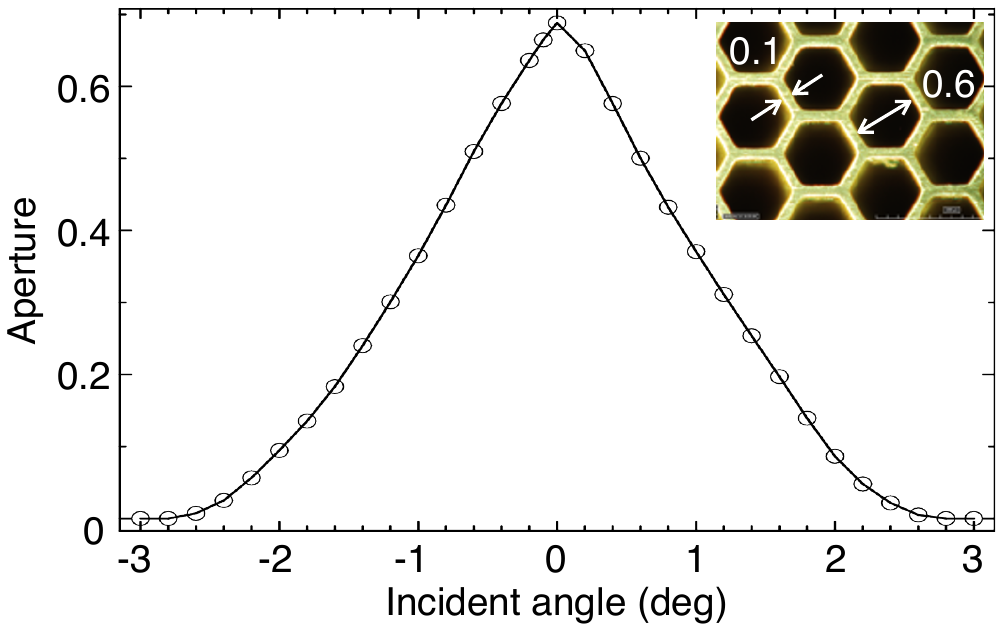}
 \end{center}
 \caption{
 The measured angular response of the collimator.
 The inset figure is a micrograph of a part of the hexagonal aperture.
 The units for the values written in the inset are mm.
 The angle scan was performed by tilting the X-ray beam in the direction indicated by the double-headed arrow of the inset figure.
 {Alt text: A graph with an inset graph.}
 }
 \label{fig:gmc_collimator_response}
\end{figure}

The collimator features a structure of hexagonal apertures, having a distance across flats of 600~$\mu$m and a depth of 15~mm.
It was assembled by stacking 300 sheets of 50~$\mu$m-thick etched austenitic stainless steel (Type 304) and fixing them together using a diffusion bonding technique.
The collimator defines the FoV of the GMC, being approximately $2^{\circ}\hspace{-1.0mm}.1$ (full-width at half-maximum; FWHM). 
The measured aperture fraction of the collimator is shown in figure~\ref{fig:gmc_collimator_response}.
A 100~$\mu$m bridge exists between the hexagonal apertures as shown in the inset micrograph of figure~\ref{fig:gmc_collimator_response}, resulting in an opening fraction of 69\% for X-ray point sources entering from the front.

\subsubsection{X-ray sensor}

\begin{figure}
 \begin{center}
  \includegraphics[width=80mm]{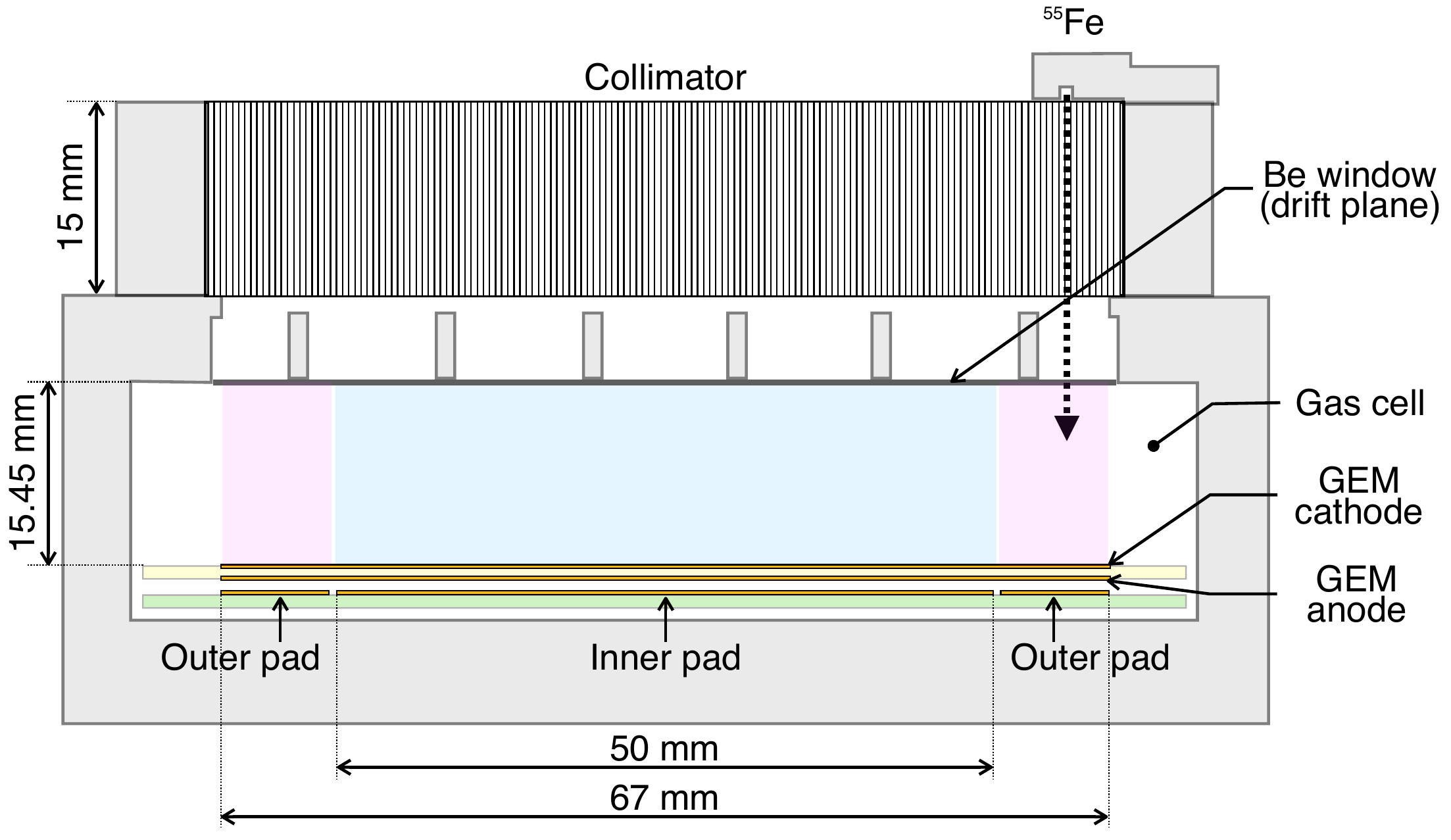}
 \end{center}
 \caption{
 A schematic cross-section of the collimator and gas cell of the GMC.
 The regions from which X-ray signals are collected by the inner and outer pads are indicated with blue and red hatching, respectively.
 {Alt text: A drawing.}
 }
 \label{fig:gmc_cross_section}
\end{figure}

\begin{figure}
 \begin{center}
  \includegraphics[width=80mm]{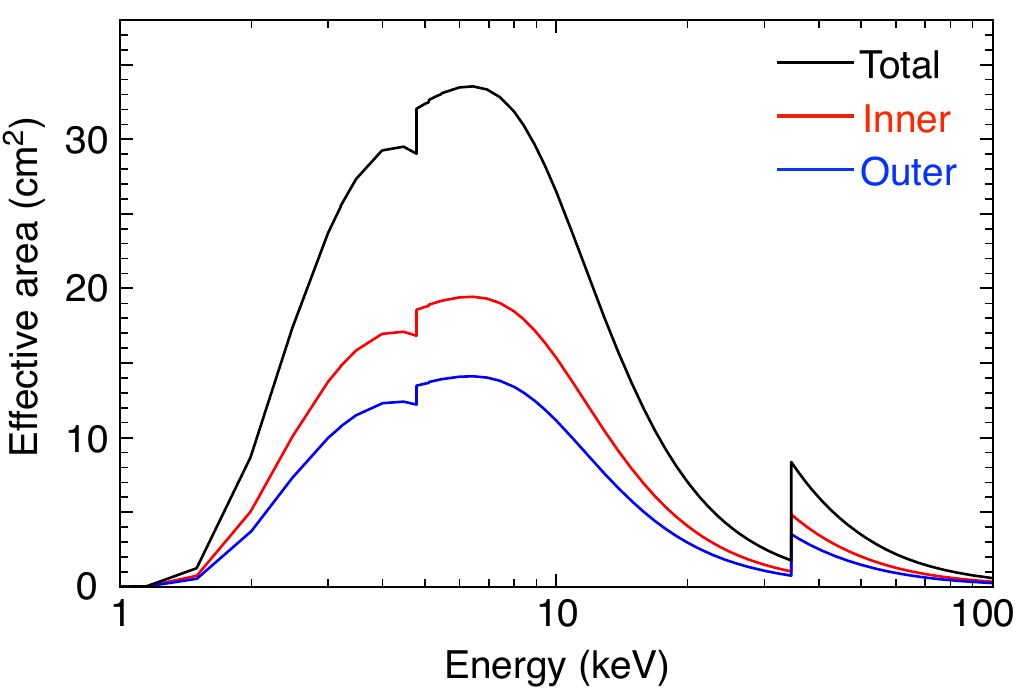}
 \end{center}
 \caption{The effective area combining GMC1 and GMC2.
 The value was derived from simulations incorporating actual measurements and experimental data.
 {Alt text: A graph.}
 }
 \label{fig:gmc_effective_area}
\end{figure}

The X-ray sensor of the GMC is a Xe-based gas proportional counter. 
An X-ray entering the gas cell through a 100-$\mu$m beryllium (Be) window is converted into a photoelectron via photoelectric absorption, followed by gas ionization as the photoelectron loses its kinetic energy.
The electrons generated by the ionization drift toward the electrodes at the bottom of the gas cell.
Before reaching the electrodes, the electrons are multiplied by a Gas Electron Multiplier (GEM, \cite{tama2009}) shown in figure~\ref{fig:nsat_gmc}b.
Figure~\ref{fig:gmc_cross_section} shows a cross-section of the collimator and the gas cell.
A voltage of approximately 600~V is applied between the GEM anode and cathode.
The GEM amplifies the signal electrons by about 500 times.
The values of the applied voltages are summarized in table~\ref{tab:gmc_fact}. 
The gas mixture consists of Xe, Ar, and dimethyl ether (DME) in a volume ratio of 75\%, 24\%, and 1\%, respectively, at a pressure of 1.2~atm at 0$^{\circ}$C.
The height of the gas target region is 15.45~mm.
The selection of gas mixture is detailed in \citet{takeda_mpgd2022}.
Figure~\ref{fig:gmc_effective_area} shows the effective area of two GMCs.

The readout electrode of the GMC consists of two pads (figure~\ref{fig:nsat_gmc}c): a circular inner pad with a diameter of 50~mm and an annular outer pad with an outer diameter of 67~mm, which is concentric with and separated from the inner pad by a 0.1~mm spacing.
For energy calibration in orbit, an $^{55}$Fe radioactive source is attached to the edge of the collimator (figure~\ref{fig:gmc_cross_section}), continuously irradiating a portion of the outer pad with 5.9~keV X-rays.
The intensity of $^{55}$Fe was approximately 1 count per sec (cps) at the time of the launch.

To prevent the diffuse X-ray background (Cosmic X-ray background; CXB), Sn and Mo shields are installed from the outside in this order. 
These shields primarily block the high-energy CXB photons.
An 800~$\mu$m-thick Sn and a 50~$\mu$m-thick Mo shields are used on the upper surface (excluding the opening fraction) and the surrounding areas of the gas cell. 
On the lower surface of the gas cell, a 500~$\mu$m-thick Sn and a 50~$\mu$m-thick Mo shields are employed.
To further prevent the intrusion of the CXB photons through gaps, an additional Sn shield of 500~$\mu$m-thick is installed on the lower surface of the aluminum housing of the FEC, and a 500~$\mu$m-thick Sn shield is installed on the sides of the housing.
Figure~\ref{fig:nsat_gmc}d shows an overall photograph of the GMC.

In recent astronomical X-ray observations, semiconductor detectors are primarily used. 
However, a gas X-ray detector was chosen for several reasons. 
A gas detector can enlarge the effective area at less expense. 
Compared to semiconductor detectors, gas detectors consume less power and generate minimal heat, making them an ideal choice for resource-constrained satellites like CubeSats.

\subsubsection{Electrical boards}

The signal processing of the GMC is performed with an analog signal processing board, the FEC, and a digital signal processing board, the DAQ.
In addition to analog signal processing, the FEC is equipped with a circuit for applying high voltage (HV) to the X-ray sensor (figure~\ref{fig:gmc_block_diagram}).
The HV component UMHV0520 (HVM Technology) is employed, and an aluminum housing is added to prevent switching noise.
The output voltage is stepped down using three 100~M$\Omega$ resistors supplied to the sensor.
A 470~k$\Omega$ resistor is serially connected at the end of the resistor chain, and the voltage drop in the resistor is used for monitoring HV value.
The voltages applied to each electrode are summarized in table~\ref{tab:gmc_fact}.

The signal from the readout pad is fed into the charge sensitive preamplifier A225 (Amptek) through a 470~pF HV capacitor on the FEC board.
The amplified signals are sent to the main amplifier on the DAQ board, which allows the signal gain to be adjusted via command.
After digitization with a 25~MHz analog-to-digital converter (ADC), the waveform data is sent to FPGA (Advanced Micro Devices Spartan 6) on the DAQ board.
The FPGA generates a trigger when a signal exceeds a preset threshold and assigns a timestamp simultaneously with an internal time counter.
These data are then sent to the STM32H7 Micro Control Unit (MCU), where digital waveform processing is performed by software.
The processed information of the signal, including peak height and rise time, is sent to the PC.
A test pulse can be sent from the digital-to-analog converter (DAC) to the A225 preamplifier through a 1~pF capacitor to test the circuit.
The maximum signal processing capability is approximately 800~Hz per GMC.

\subsubsection{Operation}

\begin{figure}
 \begin{center}
  \includegraphics[width=80mm]{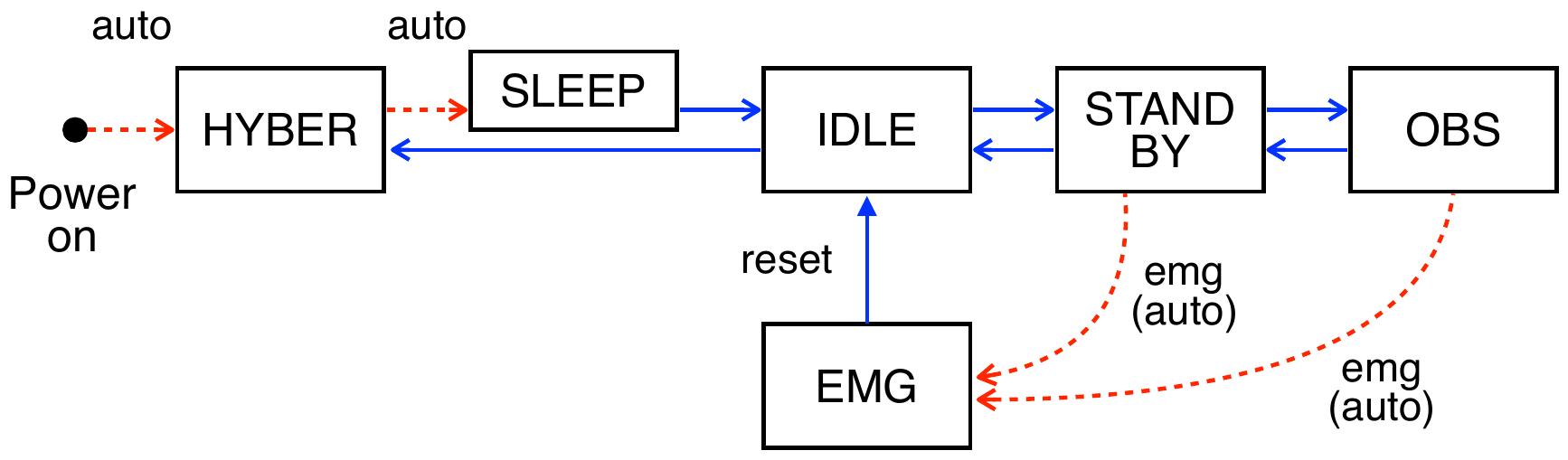}
 \end{center}
 \caption{A diagram of the GMC state machine. 
 The state transition is mainly triggered by pre-scheduled commands shown in the blue solid arrows but is autonomous at power-on and emergency (emg) cases shown in the red dashed arrows.
 Astronomical observations are conducted in the OBS state. 
 In the normal operations, the GMC goes back and forth between the IDLE and OBS states.
 The detail of the states is described in the text.
 {Alt text: A diagram.}
 }
 \label{fig:gmc_state_machine}
\end{figure}

The onboard software of the GMC is designed as a state machine.
Figure~\ref{fig:gmc_state_machine} shows a diagram illustrating the states the GMC can take and their interrelationships. 
Commands executable in each state are predefined to prevent misoperation.
Immediately after the GMC is powered on, the GMC state transitions automatically from HYBER, where only the DAQ board is powered on, to SLEEP, where the FEC is also powered on.
In the SLEEP state, commands related to high voltage cannot be issued, but all GMC housekeeping data can be retrieved.

The three states, IDLE, STANDBY, and OBS, are used in regular operations. 
The HV module is turned on in the IDLE state, but the output voltage is set to 0~V. 
In the STANDBY state, a voltage sufficiently low to prevent gas amplification in the GMC is applied.
An operation HV is applied in the OBS state, allowing astronomical observations.
During S-band communication with a ground station, the GMC remains in the IDLE state to avoid noise contamination from the S-band transmitter.
When passing through the South Atlantic Anomaly (SAA) and auroral zones surrounding the North and South Poles, where the flux of energetic particles increases, the GMC is set to either the IDLE or STANDBY state.

The GMC is operated via commands sent from the PC through the CAN bus 2. 
The protocol used is the CubeSat Space Protocol (CSP).
Commands uploaded from a ground station to the satellite are executed according to a pre-scheduled command script. In this process, the PC sends commands to the GMC at the scheduled time for execution on the DAQ board.
The GMC firmware for FPGA and MCU can be updated in orbit to fix any software bugs discovered after the launch.
The firmware uploaded to the PC is sent to the DAQ board via UART lines and updated upon receiving a boot signal from the PC.

\subsection{Radiation belt monitor (RBM)}

\begin{figure}
 \begin{center}
  \includegraphics[width=80mm]{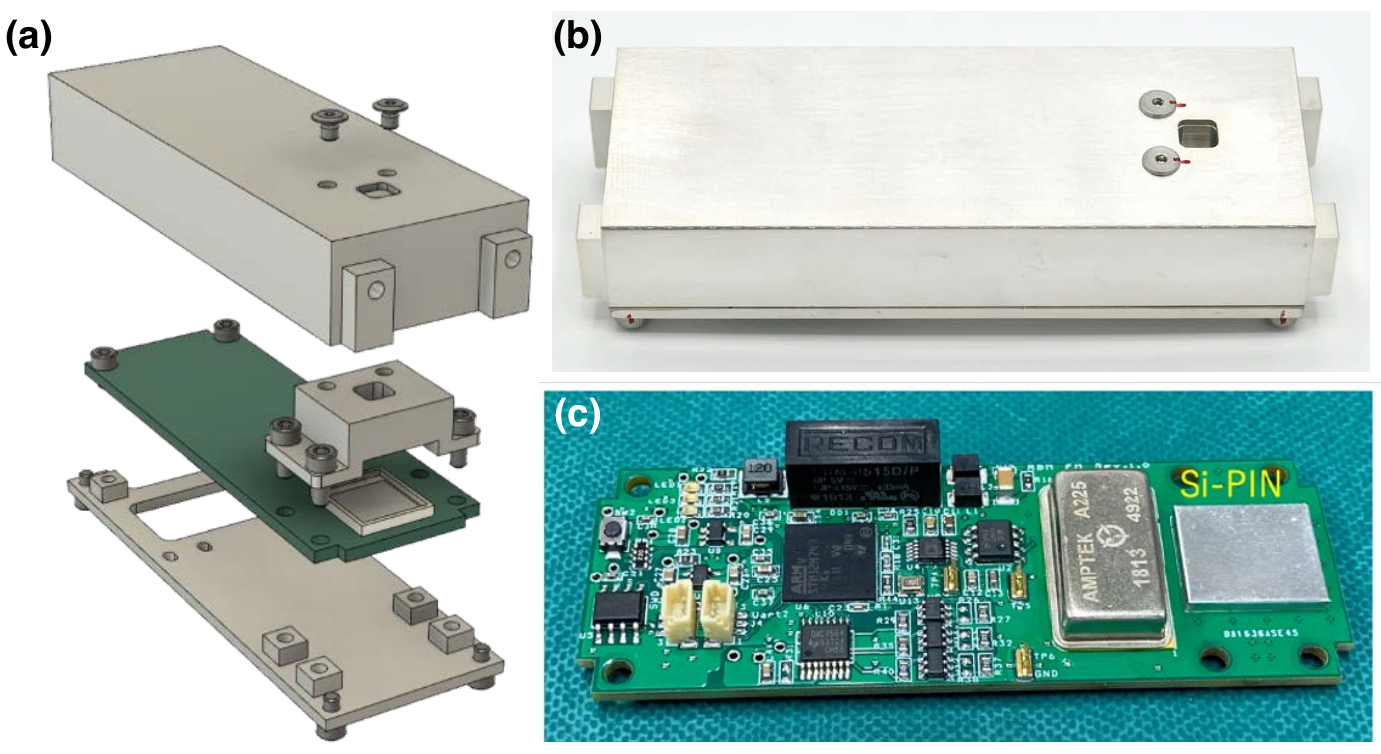}
 \end{center}
 \caption{
 (a) An exploded view of the RBM, which consists of a cover, a collimator, an RBM board, and a base from top to bottom.
 (b) A photograph of a flight model RBM. 
 The size is 95~mm$\times$32~mm$\times$17~mm. 
 (c) A photograph of the RBM board.
 {Alt text: A graph consists of one illustration and two photographs.}
 }
 \label{fig:nsat_rbm}
\end{figure}

RBM monitors the flux of charged particles (mainly protons and electrons) in the orbit of NinjaSat and issues an alert to the GMC when the flux rises above a preset threshold \citep{Kato2023SSC}.
Figure~\ref{fig:nsat_rbm}a shows an exploded view of the RBM, and figure~\ref{fig:nsat_rbm}b shows a photograph of the RBM.
A Si-PIN diode with a sensitive area of $9\times9$~mm$^2$ and a thickness of 500~$\mu$m is used as a sensor. 
A bias voltage of 30~V is applied between the anode and cathode of the Si-PIN, forming a depletion layer with a thickness of about 300~$\mu$m.
There is an aluminum collimator with a 5$\times$5~mm$^2$ hole and a thickness of 7~mm, which prevents the entry of low-energy electrons from outside the field of view. 
The RBM board, shown in figure~\ref{fig:nsat_rbm}c, is equipped with the same STM32H7 MCU as the DAQ board of the GMC and can communicate with the payload controller and the GMC using the CAN bus. 

A 100~$\mu$m thick aluminum foil is installed in front of the Si-PIN diode to block optical light. 
Due to this aluminum and the insensitive layer of the Si-PIN diode, the minimum detectable energy by the RBM is limited to approximately 200~keV for electrons and 5~MeV for protons. 
The RBM is equipped with three comparators to discriminate events by energy deposit in the diode based on signal amplitude.
Each comparator output is connected to a counter (counter 0, 1, or 2) that counts the number of signals passing through the comparator. 
The three comparators are set so that the energy loss in the Si-PIN is approximately 100, 150, and 300~keV.
Alerts are issued using only Counter 2, which corresponds to a proton kinetic energy of approximately less than 200~MeV.
The RBM evaluates the counter values every second.
If the value of Counter 2 exceeds 12~cps, an alert is sent to the GMC via the CAN bus 1.
The detailed RBM specifications are summarized in table~\ref{tab:rbm_fact}.

\begin{table}
  \tbl{RBM parameters and specifications.}{%
  \begin{tabular}{lr}
      \hline
      Parameter & Value or note\\ 
      \hline
      Size & $95\times32\times17$~mm$^3$\\
      Weight & 70~g\\
      Operational temperature & $-10^{\circ}$C to $+30^{\circ}$C\\
      Power consumption & 1~W (maximum)\\
      Voltage & +5~V (regulated)\\
      Al collimator FoV & 45$^{\circ}\times$45$^{\circ}$ \footnotemark[$*$]\\
      Si-PIN diode&\\
      \hspace{5mm} Chip size & 9$\times$9~mm$^2$\\
      \hspace{5mm} Chip thickness & 500~$\mu$m\footnotemark[$\dag$]\\
      \hspace{5mm} Bias voltage & 30~V\\
      Threshold kinetic energy & \\
      \hspace{5mm} Electron & $\ge200$~keV\\
      \hspace{5mm} Proton & $\ge5$~MeV\\
      Comparator threshold \footnotemark[$\ddag$]& \\
      \hspace{5mm} Counter 0 & $\ge100$~keV\\
      \hspace{5mm} Counter 1 & $\ge150$~keV\\
      \hspace{5mm} Counter 2 & $\ge300$~keV\\
      Alert threshold & $\ge12$ cps in Counter 2\\
      \hline
      \end{tabular}}\label{tab:rbm_fact}
\begin{tabnote}
\footnotemark[$*$] Only effective for low energy electrons.\\ 
\footnotemark[$\dag$] The thickness of depletion layer is approximately 300~$\mu$m. \\ 
\footnotemark[$\ddag$] 
The threshold is applied to the energy deposited in the Si-PIN sensor.\\ 
%\footnotemark[$\|$]  ... \\
%\footnotemark[$\sharp$]  ... \\  
%\footnotemark[$**$]  ... \\ 
%\footnotemark[$\dag\dag$]  ... \\ 
\end{tabnote}
\end{table}

\subsection{Safety mechanism for GMC operation}

Given the extensive use of commercial off-the-shelf components, radiation tolerance of the circuit parts is particularly crucial. 
Every active component used in payloads has undergone individual radiation testing at the Heavy Ion Medical Accelerator in Chiba (HIMAC) and the Wakasa Wan Energy Research Center (WERC) to ensure they can withstand more than two years of operation. 
However, there is always a risk of latch-up caused by cosmic rays in FPGA and MCU. 
If the GMC hangs up, it cannot receive the command to turn off the HV, potentially leading to excessive discharge when entering a high-radiation region.
A ping watchdog mechanism is implemented in the EPS to mitigate this risk. 
The feature sends a ping to the GMCs every 20~s, and if no response is received from the GMC three consecutive times, the power to the GMC is forcibly shut down.
This mechanism ensures that the GMC can be powered down within a maximum of 60~s under critical circumstances. 
We can adjust parameters such as ping intervals and the maximum allowable number of attempts via command.

During normal operations, the GMC transitions state according to commands at a predetermined date and time.
If, for some reason, the state transition command fails and HV remains applied, an automatic HV shutdown feature activates when the count rate of the GMC exceeds a preset threshold (1.2$\times$10$^4$~counts per 8~s). 
This function is effective only in the STBY and OBS states, which involve high voltage applications, transitioning to the emergency (EMG) state illustrated in figure~\ref{fig:gmc_state_machine}. 
In the EMG state, no commands related to the HV operation are accepted, and the system returns to normal operational status only after explicitly receiving a reset command from the ground.
Consequently, the GMC incorporates a four-fold safety mechanism consisting of (1) HV control via nominal operation commands, (2) a self-count limiter on count rate, (3) RBM alerts via the CAN bus, and (4) the ping watchdog by the EPS.

\subsection{Payload environment tests and calibrations}

\begin{figure*}
 \begin{center}
  \includegraphics[width=160mm]{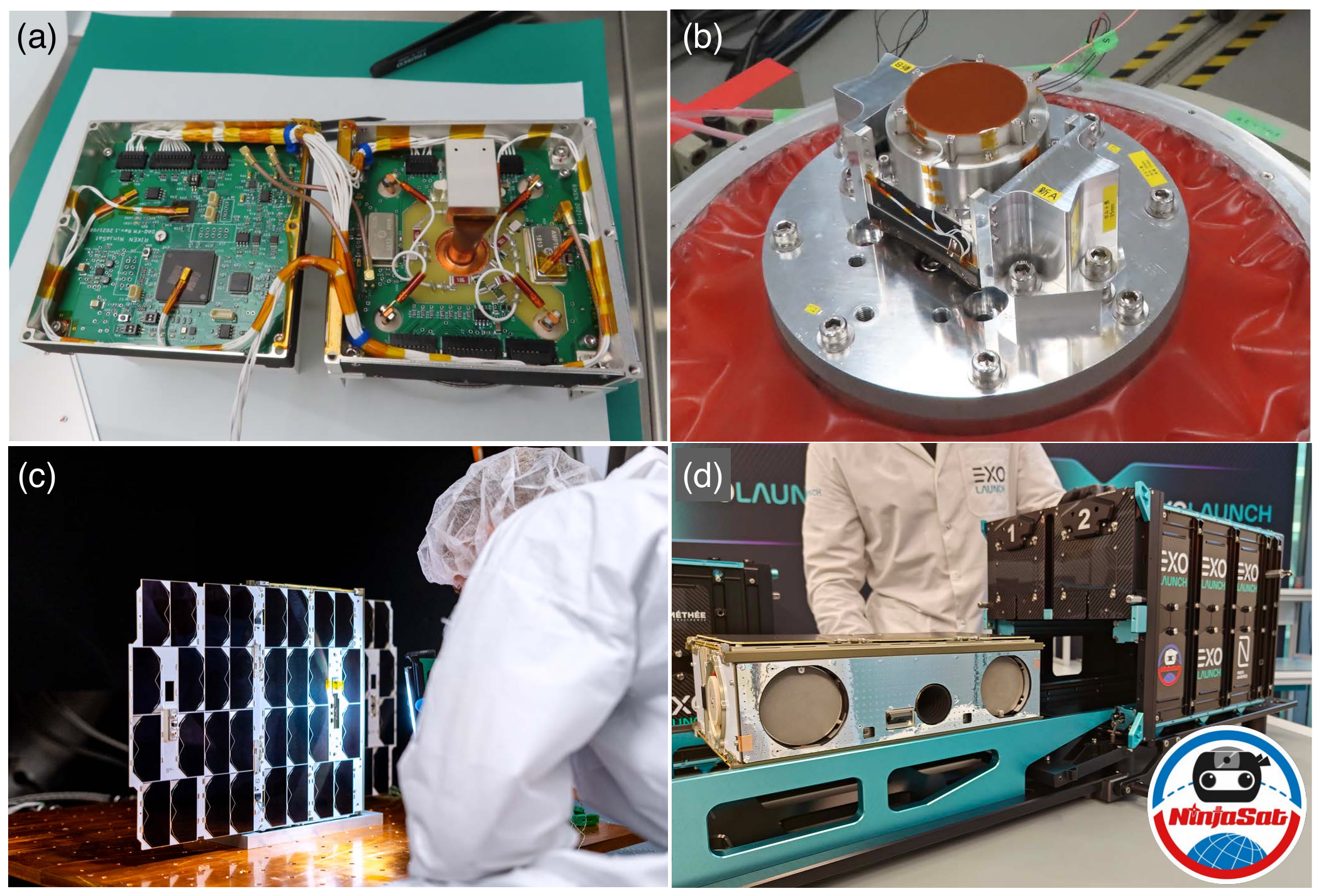}
 \end{center}
 \caption{
 Snapshots taken during the Ninjasat development.
 (a) The electrical boards photographed during GMC2 assembly.
 The right board is the FEC, and the left one is the DAQ.
 Harnesses between the FEC and DAQ are almost assembled.
 (b) The GMC on the vibration test bench.
 (c) NinjaSat in the thermal vacuum chamber.
 (d) NinjaSat stowing into the EXOpod deployer.
 The inset at the bottom right corner is the NinjaSat emblem.
 {Alt text: A graph consists of four photographs.}
 }
 \label{fig:nsat_photos}
\end{figure*}

The flight models of the GMC and RBM, which were assembled at RIKEN, underwent vibration testing at the Saitama Industrial Technology Center. 
Some photographs are shown in figures~\ref{fig:nsat_photos}a and \ref{fig:nsat_photos}b.
We applied the random vibration test profile applicable to CubeSats mounted on the SpaceX Falcon 9.
The test levels were 7.07~G$_{\rm rms}$ and 5.13~G$_{\rm rms}$ along each axis for GMC1 and GMC2, respectively, with a duration of 60~s.
The higher vibration level for GMC1 was due to the uncertainty of the launch vehicle at the time of testing; thus, a precautionary +3dB vibration test, equivalent to a proto-flight test (PFT), was applied. 
Both RBM1 and RBM2 underwent a random vibration test at a PFT level of Falcon 9, specified as 7.10~G$_{\rm rms}$ for 60~s. 

\begin{figure*}
 \begin{center}
  \includegraphics[width=160mm]{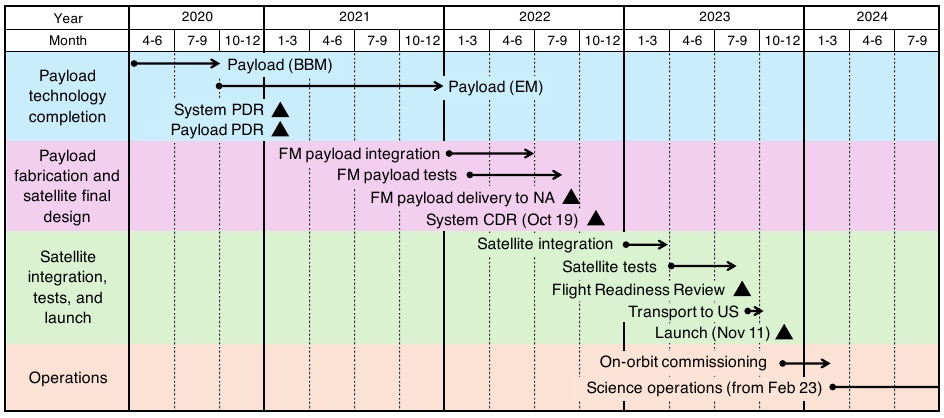}
 \end{center}
 \caption{
 The NinjaSat project timeline.
 Acronyms are BBM (breadboard model), EM (engineering model), PDR (preliminary design review), FM (flight model), NA (NanoAvionics), and CDR (critical design review).
 {Alt text: A chart.}
 }
 \label{fig:nsat_timeline}
\end{figure*}

We conducted thermal vacuum tests of the payloads using the thermal vacuum chamber at the Institute of Space and Astronautical Science, Japan Aerospace Exploration Agency (ISAS/JAXA).
Each payload underwent four cycles of thermal vacuum ranging between $-30^{\circ}$C and $+60^{\circ}$C, with a temperature change rate of 0.6$^{\circ}$C~min$^{-1}$. 
In one of these cycles, we conducted performance evaluation and functional tests of all payloads between $-10^{\circ}$C and $+30^{\circ}$C with a margin of $\pm10^{\circ}$C around the intended operational temperature range (0--20$^{\circ}$C).
Additionally, we performed turn-on tests at $-20^{\circ}$C and $+40^{\circ}$C to confirm that each payload was powered on at these temperatures without any troubles.

We evaluated the performance of the GMCs using an X-ray generator at RIKEN and the Photon Factory (PF) BL-14A beamline at the High Energy Accelerator Research Organization (KEK). 
At RIKEN, characteristic X-rays of 4.5, 6.4, 8.0, and 17.4~keV from an X-ray generator were used to confirm the linearity of the detector response with respect to the incident energy. 
The detection efficiencies of the GMCs were measured at the same energies by performing a simultaneous comparison with Si-PIN and CdTe semiconductor detectors.
During the thermal vacuum tests, we found that the electrical gain of the GEM exhibited temperature dependence.
To investigate and model the temperature dependence, we performed a gain scan of the entire sensitive area of the GMCs with a small X-ray generator that provides 4.5~keV X-rays inside a temperature-controlled chamber.
The data was taken in the temperature range between $-10^{\circ}$C and $+27^{\circ}$C for the GMC1 and between $-10^{\circ}$C and $+24^{\circ}$C for the GMC2.

At KEK-PF, we irradiated X-rays onto the GMCs over an energy range of 6.4 to 50~keV to confirm the linearity of energy response over a wide energy range, to assess charge sharing near the boundary between the inner and outer electrodes, and to observe changes in energy response around the Xe K-edge at 34.56~keV. 
Additionally, the detection efficiencies of the GMCs were measured above 6.4~keV.
We evaluated the performance of the RBMs at WERC by irradiating them with a 100~MeV proton beam.
We also performed the calibration of the deposited energy in the RBMs at RIKEN by using the 60~keV X-rays from the radioactive source $^{241}$Am.

\subsection{Payload integration to the satellite and tests}

Two GMCs and two RBMs were delivered to NanoAvionics in 2022 August.
The electrical and communication connections were tested using Flatsat, a satellite hardware simulator, prior to payload integration into the satellite.
The satellite integration, environmental testing, and satellite quality assurance testing were conducted by NanoAvionics until July 2023. 
The vibration test was conducted using the profile recommended for the Falcon 9 rideshare, with each axis tested at 5.57~G$_{\rm rms}$ for 60~s. 
This increase from the 5.13~G$_{\rm rms}$ level used in the payload test reflects an update to the Falcon 9 User’s Manual, which raised the specified vibration level. 
Electrical tests were performed on each payload before and after the vibration test, and all payloads passed without any trouble. 

The thermal vacuum test (figure~\ref{fig:nsat_photos}c) consisted of four temperature cycles between $-20^{\circ}$C and $+50^{\circ}$C, with a temperature change rate of $\ge 1^{\circ}$C~min$^{-1}$ at a pressure of $\le$10$^{-4}$~Pa. 
Before the thermal cycle, we conducted the functional verification tests of the payloads in both hot ($+30^{\circ}$C) and cold ($-10^{\circ}$C) conditions at a pressure of $\le$10$^{-4}$~Pa to validate stable operations in flight.
To evaluate the noise environment, the GMCs and RBMs were also tested with all satellite subsystems operating.

The Flight Readiness Review (FRR) was conducted on 2023 August 14, immediately after the satellite quality assurance tests were completed and the satellite was ready for shipping. 
After the completion of the FRR, the satellite was stowed in the EXOpod (figure~\ref{fig:nsat_photos}d), the CubeSat deployer provided by Exolaunch, and transported to California, USA, in 2023 September.
A Gantt chart illustrating the project progress is presented in figure~\ref{fig:nsat_timeline}.

\section{Launch and payload verification}

\subsection{Launch and initial operation}

NinjaSat was launched at 10:49 a.m. in Pacific Standard Time (PST) on 2023 November 11 by a SpaceX Falcon 9 rocket from Vandenberg Space Force Base. 
The rocket carried 90 small satellites, including NinjaSat, as the Transporter-9 ride-share mission. 
NinjaSat was deployed into a Sun-synchronous orbit as the second of 90 satellites at 11:43 a.m. PST. 
The orbital inclination is 97$^{\circ}$ with an altitude of about 530~km, and Local Time at Descending Node is 10:32 a.m. 
The EPS of NinjaSat was turned on 30~s after the deployment as scheduled, and the first solar panel was deployed at 15:51 PST, followed by the deployment of the second solar panel at 16:04 PST.

The first contact with the satellite was established in UHF on November 14 at 19:22 UTC, three days after the launch.
The telemetry data obtained at this contact confirmed the activation of the EPS and the deployment of the solar panels, as mentioned above. 
Subsequently, satellite commissioning proceeded smoothly using both UHF and S-band communications.

\begin{figure}
 \begin{center}
  \includegraphics[width=80mm]{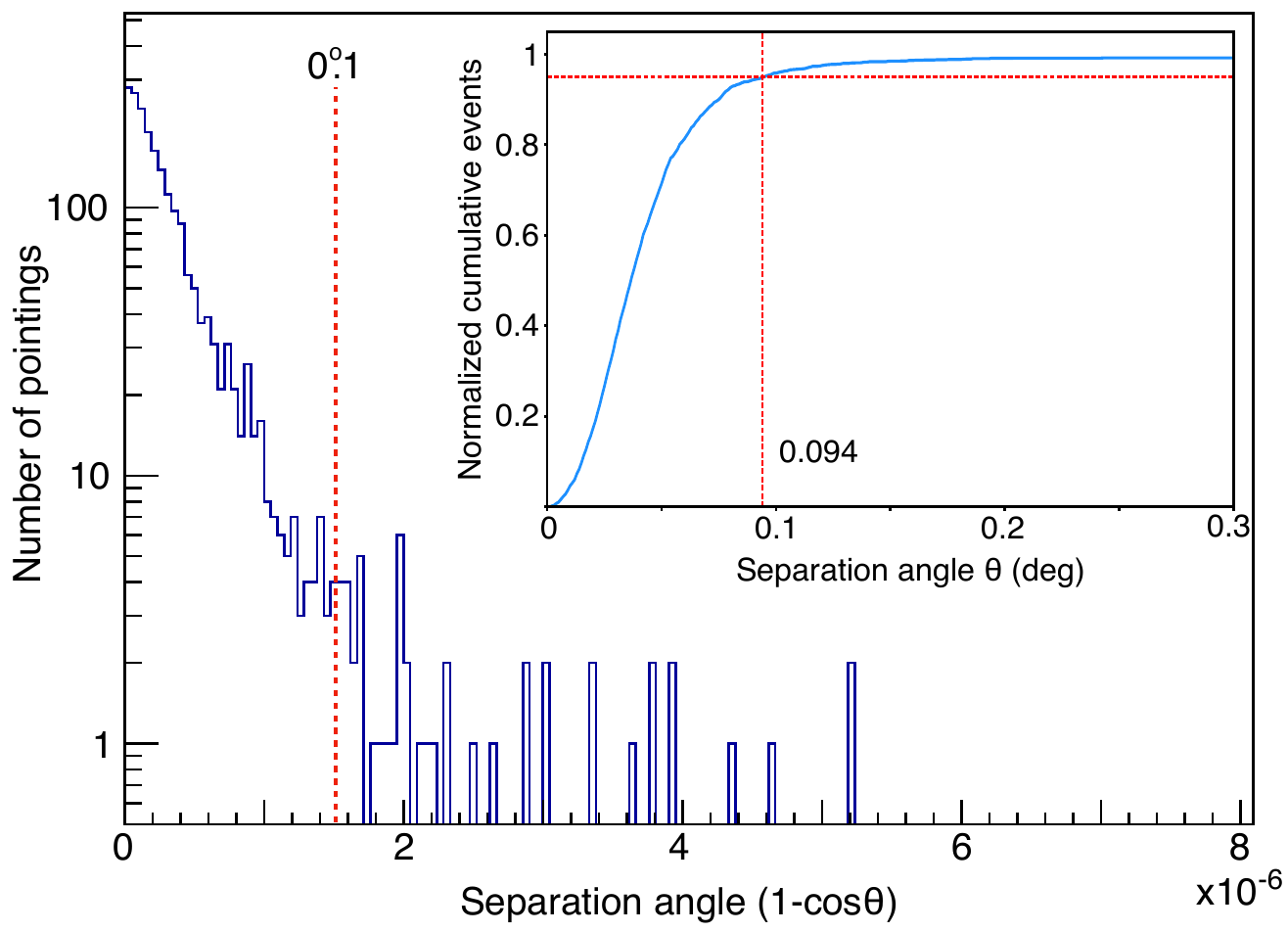}
 \end{center}
 \caption{The pointing accuracy of NinjaSat derived from the ADCS output.
  The horizontal axis measures (1-$\cos\theta$), where
 $\theta$ is the separation angle between the target coordinate and the boresight (+X in figure~\ref{fig:nsat_overview}).
The pointing data recorded every 15~s during astronomical observations and every 60~s at other times was downloaded.
 The separation angles $\theta=0^{\circ}\hspace{-1.0mm}.1$ is shown as a dotted line.
 The inset figure shows normalized cumulative events as a function of the separation angle. 
 The 95\% of the events are contained within $0^{\circ}\hspace{-1.0mm}.094$.
 {Alt text: A graph with an inset graph.}
 }
 \label{fig:nsat_pointing}
\end{figure}

The ADCS verification began on December 1, and three-axis attitude control was initiated as initially planned.
Figure~\ref{fig:nsat_pointing} shows the actual pointing accuracy when observing a celestial object from 2024 April 24 to May 6.
The pointing accuracy presented here is the angular separation between the coordinates of the celestial target and the actual line of sight direction as determined by the ADCS (+X in figure~\ref{fig:nsat_overview}).
The required specification of $0^{\circ}\hspace{-1mm}.1$ (2$\sigma$ confidence level; CL) accuracy has been sufficiently met, achieving $0^{\circ}\hspace{-1mm}.094$.
The pointing error is damped and stabilized to a constant value within 30~s after the maneuver ends.

\subsection{Payload verification}

\subsubsection{Payload startup}

On 2024 January 22, we started the payload verification.
Initially, the RBM1 and RBM2 were powered on, and then functional tests were performed, followed by a signal threshold scan to determine the noise floor of the circuit. 
The scan results were used to set the thresholds for the three comparators, which define the three energy bands of the RBM. 
The lowest energy threshold of the first counter (Counter 0) is set just above the circuit noise floor (100~keV), and it is used to monitor variations in the circuit noise condition in orbit. 
The second counter (Counter 1) is configured to be sensitive to the energy deposit by particles above 150~keV, and the third counter (Counter 2) is sensitive to the energy deposit above 300~keV.

\begin{figure*}
 \begin{center}
  \includegraphics[width=160mm]{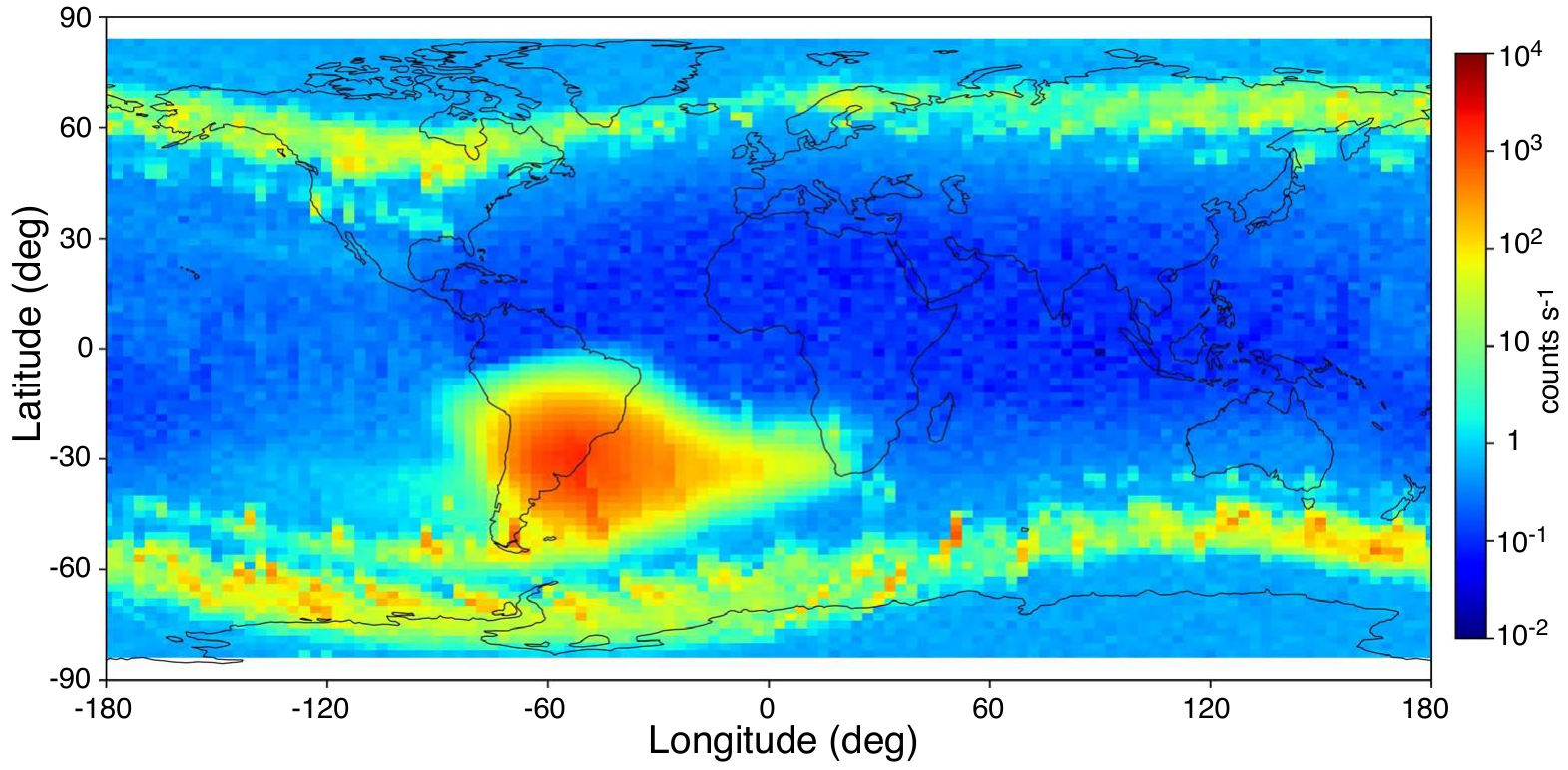}
 \end{center}
 \caption{
 A map of the particle count rate obtained with the Counter 2 of the RBM1.
 The data from 2024 February 22 to March 15 was averaged.
 The bin size is 3$^{\circ}$ (longitude) $\times$ 2$^{\circ}$ (latitude).
 %degrees in longitude and latitude, respectively.
 The South Atlantic Anomaly on the east coast of South America and auroral bands surrounding the North and South Poles are clearly seen. 
 }
 \label{fig:nsat_rbm_countmap}
\end{figure*}

Figure~\ref{fig:nsat_rbm_countmap} shows the averaged particle count rate taken with the Counter 2 of the RBM1 from February 22 to March 15 overlaid on a world map.
The SAA and the auroral zones encircling the Arctic and Antarctic are clearly visible.
The orbit of NinjaSat allows us to obtain a map of particle count rate throughout the globe approximately every six days.
We use this map to determine where the GMC can operate in orbit.

GMC1 commissioning began on January 29. 
After verifying each function except the HV-related ones, HV tests were conducted on February 6, confirming signal acquisition correctly.
On February 9, the satellite threw to the calibration target Crab Nebula, and the X-ray signals were successfully observed.
At this point, the NinjaSat project achieved the minimum success criterion. 

GMC2 commissioning began on February 12. 
The start-up process of GMC2 proceeded as planned and finished on February 21.
Subsequently, when attempting to operate the GMC1 and GMC2 simultaneously, we found that turning on the HV for both the GMC1 and GMC2 caused noise interference, leading to unintended artificial triggers. 
This phenomenon had not been observed during ground testing. Later investigations revealed that noise was generated when signals were transmitted through the CAN bus. 
However, at this point, priority was given to initiating astronomical observations, and the operation was limited to using only one unit, either the GMC1 or GMC2.

NinjaSat started science observations on 2024 March 23.
Two days earlier, on March 21, the discovery of the new X-ray source SRGA J144459.2$-$604207 was reported on Astronomer's Telegram \citep{2024ATel16464}.
We promptly modified our initial observation plans and selected this object as the first-light observation target \citep{2024ATel16495}.
The results of one month of continuous observation of this target are compiled in a separate paper \citep{takedaSRGAJ1444}.

\subsubsection{GMC timing capability}
\label{sec:gmc_timing}

\begin{figure}
 \begin{center}
  \includegraphics[width=80mm]{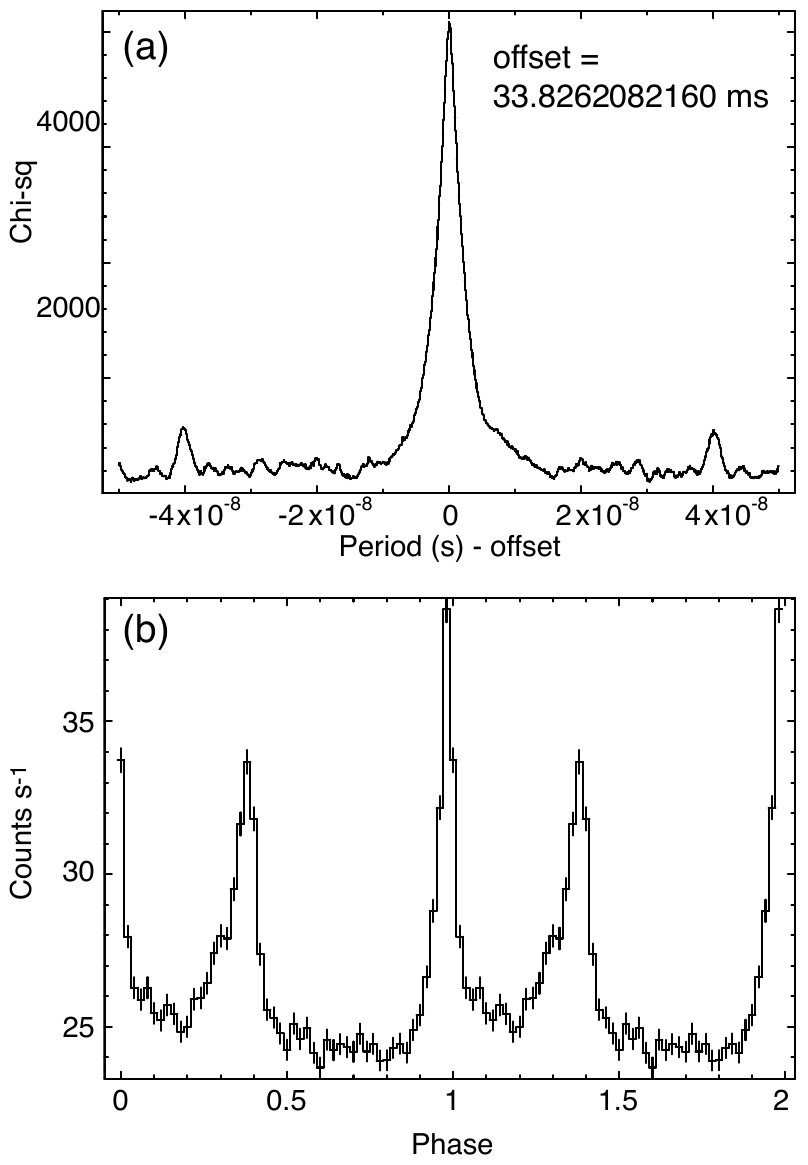}
 \end{center}
 \caption{
 (a) The X-ray periodogram of the Crab Pulsar based on observational data obtained on 2024 March 20 for 11~ks.
 The best period is shown in the figure.
 (b) The pulse profile of the Crab Pulsar, created by folding in the rotation period.
 Two rotation cycles are plotted for clarity.
 The rotation phase was assigned to each X-ray photon based on the JBO radio observations.
{Alt text: Two graphs vertically stacked.}
 }
 \label{fig:nsat_crab}
\end{figure}

Contributing to time-domain astronomy, accurately assigning the arrival time to each X-ray photon is fundamentally important for monitoring the intensity variations of X-rays. 
Figure~\ref{fig:nsat_crab}a shows a periodogram derived from X-ray data of the Crab Nebula obtained with the GMC1 for approximately 11~ks.
The data was barycentrically corrected to the solar system barycenter. 
The period of 33.8262~ms for the X-ray pulsar at the center of the Crab Nebula was successfully obtained. 
When comparing the radio observation of the Crab Pulsar by the Jodrell Bank Observatory (JBO; \cite{lyne2014JBO}), we found the difference in the pulse period was approximately 10~ns.
This confirms that at least the relative time information is properly assigned to each X-ray photon.

Using the ephemeris derived from radio observations of the Crab Pulsar provided by JBO, we assigned a rotational phase to each photon.
The pulse profile created using the phase-tagged photon data is shown in figure~\ref{fig:nsat_crab}b. 
The center of the primary peak from radio observations is placed at $phase=1$ by definition, while it is known that in X-ray observations, the primary peak appears at $phase=0.99$~\citep{terada2008, enoto2021}.
In our observations, $phase=0.99$ was obtained, confirming that the absolute time is correctly assigned with an accuracy of at least sub-milliseconds. 
We verified that the time assignment accuracy for X-ray photons meets the required specification of 1 ms.

\subsubsection{GMC spectrum and light curve}

\begin{figure}
 \begin{center}
  \includegraphics[width=80mm]{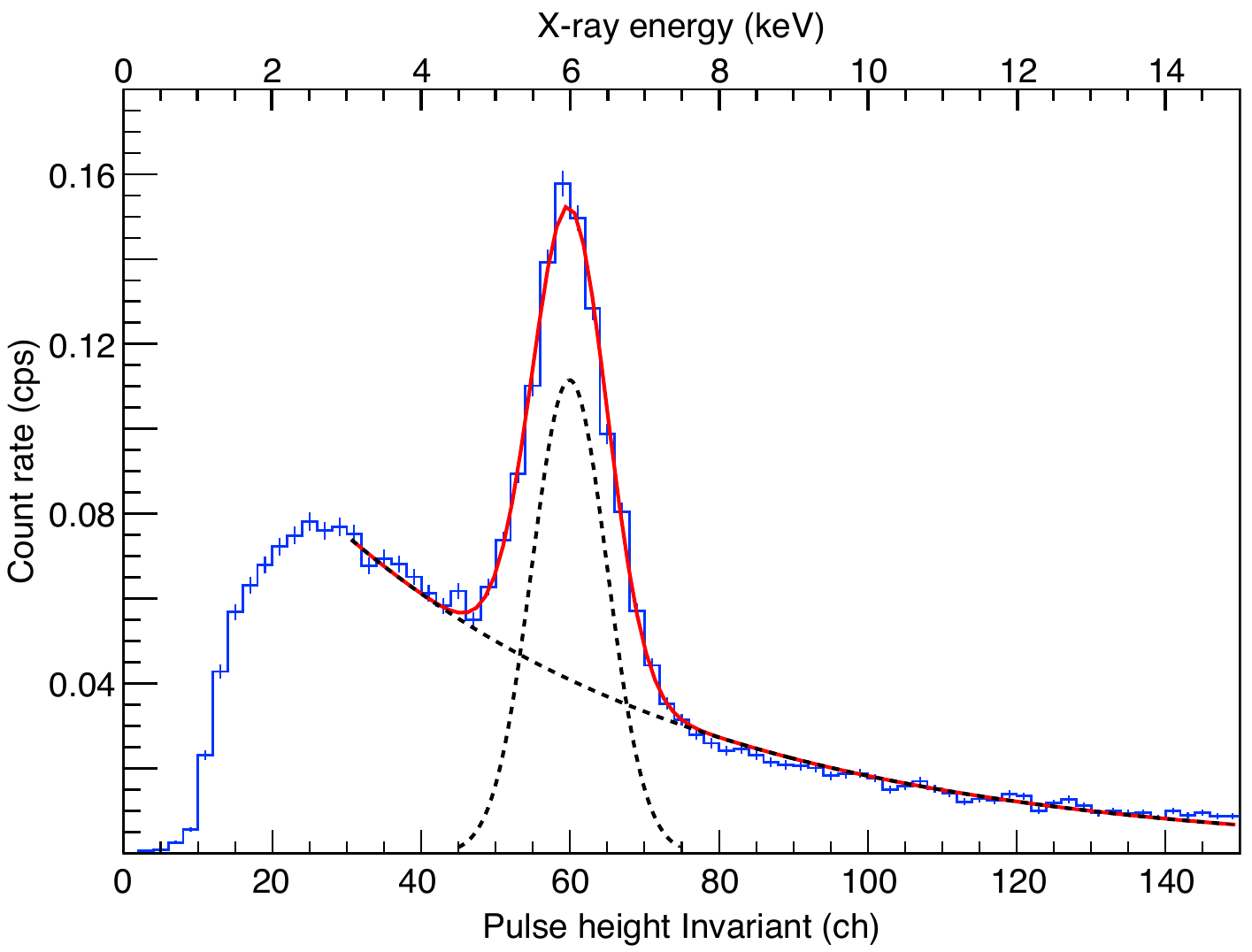}
 \end{center}
 \caption{
 The GMC1 spectrum obtained with the outer pad on 2024 March 2 and 3 with a total exposure of 16.7~ks when no X-ray sources were in the FoV.
 The conversion factor from the pulse height invariant channels to energy is approximately 0.1~keV~ch$^{-1}$.
 A distinct 5.9~keV peak from the $^{55}$Fe calibration source is seen, with an energy resolution of 19.8\% (FWHM).
 The data was fit with a model of exponential and Gaussian functions, and each component is shown with dotted lines.
 {Alt text: A graph.}
 }
 \label{fig:fe55_spectrum}
\end{figure}

Using the $^{55}$Fe calibration source mounted on top of the collimator as shown in figure~\ref{fig:gmc_cross_section}, we continuously monitor the gas gain of the GMC and energy resolution from pre-launch through post-launch. 
Figure~\ref{fig:fe55_spectrum} shows the X-ray spectrum obtained with the GMC1 outer pad over two days, from 2024 March 2 through 3. 
The peak of the 5.9~keV X-rays emitted from $^{55}$Fe is visible in the figure, with an energy resolution of 19.8\%.
The gas gain showed minimal variation ($<5$\%) over 1~yr before the launch. 
However, the gain increased by 12--13\% after the launch.
During the GMC commissioning phase on orbit, the gain increased by 7\% in the first 10~d.
After the commissioning phase, the gain increased slightly by 3\% with a time constant of approximately 90~d.
The gain currently stabilized and showed few variations, indicating likely saturation.
Although several possibilities, such as the expansion of the gas cell in space, could explain this gain increase, the exact cause remains uncertain.
The energy resolution between ground tests and space operations remains unchanged.
The X-ray count rate from the $^{55}$Fe calibration source has been confirmed to be consistent, within the range of systematic error, with the values observed during ground testing.

\begin{figure}
 \begin{center}
  \includegraphics[width=80mm]{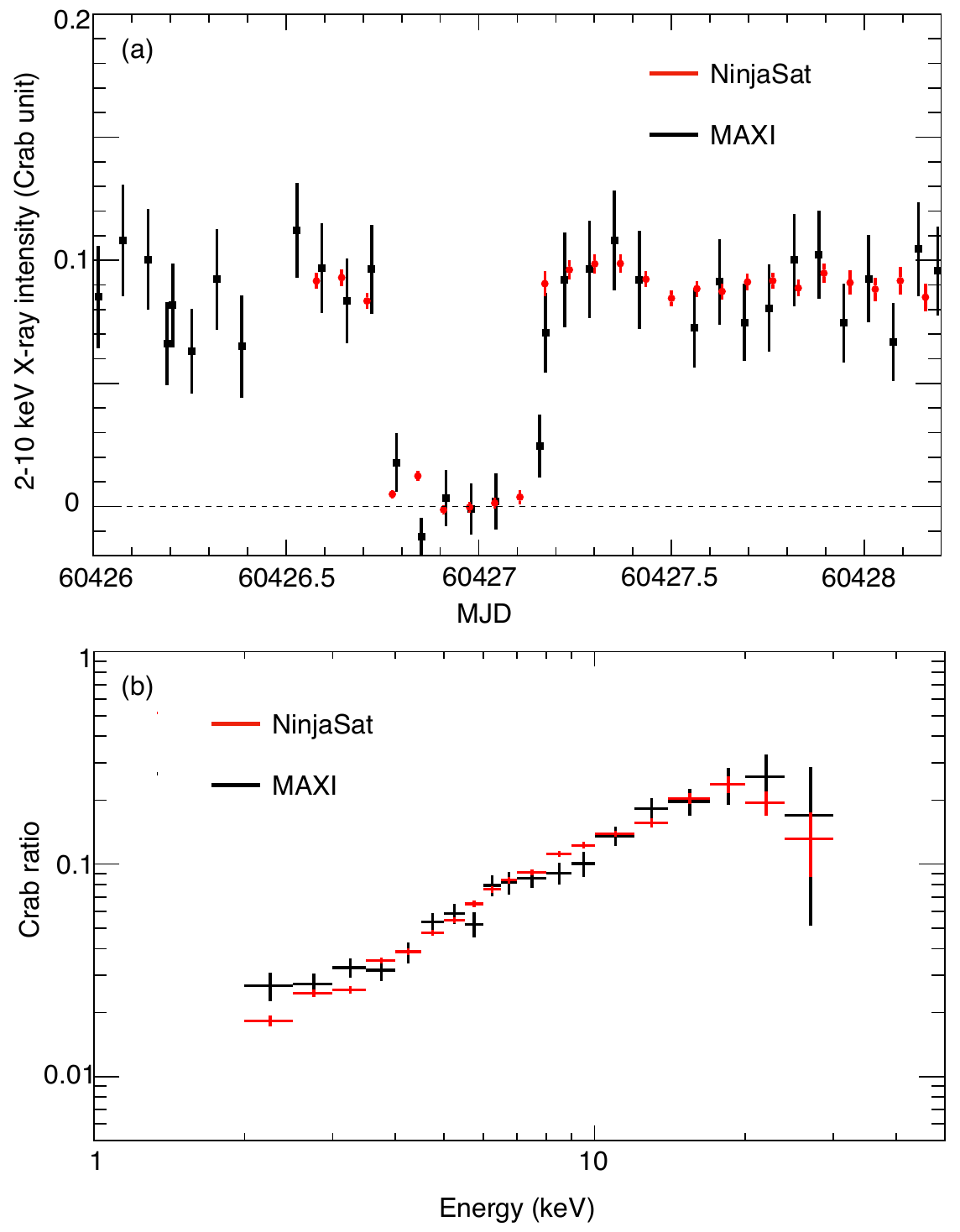}
 \end{center}
 \caption{
 (a) The 2--10~keV light curves of Her X-1 observed with the GMC1 inner pad (red) and MAXI (black) between 2024 April 26 and 28.
The exposure of each data point is typically 250--1200~s for NinjaSat and 60~s for MAXI.
 The monitoring was conducted during a main-on phase of Her X-1, and the eclipse by the companion star was recorded around MJD 60427.
 (b) The Crab ratio X-ray spectra of Her X-1 obtained with the GMC1 inner pad (red) and MAXI (black).
 The shape of the Crab ratio spectrum resembles the $\nu F\nu$ plot (spectral energy density) since the Crab nebula follows the power-law shape with its photon index of approximately 2.
The GMC1 spectrum was plotted with 71~ks data containing the period shown in (a).
{Alt text: Two graphs vertically stacked.}
 }
 \label{fig:nsat_her_x-1_lc_spec}
\end{figure}

In long-term monitoring of X-ray sources, energy-selected light curves provide crucial information on intensity variations. 
NinjaSat features the largest X-ray detection area among CubeSat missions, making it highly suitable for monitoring the X-ray intensities of celestial objects.
Figure~\ref{fig:nsat_her_x-1_lc_spec}a shows the light curve for Hercules (Her) X-1 observed by the GMC1 over a two-day period from 2024 April 26 to April 28.
Her X-1, a neutron star binary, is known to exhibit eclipses as the neutron star is periodically obscured by its companion star every 1.70~d. 
The light curve obtained from MAXI observations over the same period is overlaid to trace long-term variations. 
Intensity variations and a dip due to the eclipse are consistent across both MAXI and NinjaSat data.
Additionally, NinjaSat’s larger effective area and longer observation period, compared to those of MAXI, result in reduced statistical errors.
While MAXI observes each object for only about 60~s every 90~min, NinjaSat is capable of continuous observation (more than 1000~s per 95~min) of the same object.

The energy response of the GMC is known to depend slightly on temperature.
Evaluation of this dependence and refinement of a temperature-corrected energy response model are currently underway. 
Figure~\ref{fig:nsat_her_x-1_lc_spec}b shows the X-ray spectrum of Her X-1 divided by the spectrum of the Crab Nebula, a standard source in X-ray astronomy, producing what is known as the Crab ratio spectrum. 
This approach cancels uncertainties in the energy response, allowing for direct comparison with the Crab ratio spectrum from other satellites.
%The spectrum is averaged across both orbit and temperature.
In the same figure, we overlay the Crab ratio spectrum of Her X-1 observed around the same period by the MAXI/GSC.
Although a detailed analysis of the spectrum is beyond the scope of this paper, the obtained spectrum closely matches that of the MAXI/GSC, accurately reproducing the spectral turnover near 20~keV (e.g., \cite{enoto2008herx1, asami2014herx1}).

\subsection{Science operations}

Communication with the satellite using the S-band is primarily conducted via the KSAT Svalbard ground station.
Since NinjaSat is in a polar orbit, the high latitude of the Svalbard station provides frequent communication opportunities, making it an ideal choice. 
Additionally, the Awarua ground station in New Zealand can be used as a backup and is reserved for emergencies.
Communication between the satellite and ground stations is managed by the operators at NanoAvionics. 
Reservations for ground station passes are made online approximately three days in advance.
 
NinjaSat can observe any X-ray object as long as the target is not restricted by the Sun angle limitation of the star tracker: The Sun angle limitation (the angle between the +X axis of the satellite and the direction to the Sun) is 40$^{\circ}$.
However, when observing X-ray sources during the daytime in orbit, operations are conducted to ensure that the battery charging duty cycle exceeds 50\% of the orbit period to avoid hindering solar panel battery charging.
This restriction may be relaxed as operational understanding improves. 
The effective duty cycle is calculated by multiplying the charging time by $\cos\theta_S$, where $\theta_S$ is the angle between the normal to the solar panels (+Y in figure~\ref{fig:nsat_overview}) and the direction of the Sun as seen from the satellite. 
Since the daytime period on orbit is about 63\%, restrictions on daytime observations occur when $\theta_S >$37$^{\circ}$. 
NinjaSat, being a satellite in a polar orbit, passes through high-radiation regions near the North and South Poles, during which GMCs suspend observations to protect themselves.
These periods can be used for battery charging, giving NinjaSat enough time to recharge.
Therefore, NinjaSat can practically observe any X-ray sources without restrictions, as long as the satellite does not violate the Sun angle limitation.

The NinjaSat maneuvering speed is 5$^{\circ}$~s$^{-1}$.
Due to the small moment of inertia of NinjaSat, the time required from the end of a pointing change to stabilization is usually within one minute, based on actual performance.
This is a feature unique to small satellites, unlike larger satellites.
The ADCS information, including the pointing directions, is recorded for downlink every 60~s; however, during astronomical observations, the recording rate increases to every 15~s.

Scientific operations are conducted according to the following procedure. 
First, the NinjaSat team creates the list of target objects to be observed, which contains the start and stop time of the observation and the target coordinate (RA, Dec)$_{\rm J2000.0}$ in the equatorial coordinate system.
During this process, the solar separation angle of the objects and the charging duty cycle are considered. 
High-radiation regions where the GMC cannot operate are pre-mapped, and if they coincide with daytime on orbit, they are allocated for charging. 
After the ADCS team at NanoAvionics confirms the pointings are safe for the satellite, the attitude file is uplinked to the satellite.
Second, the NinjaSat team prepares the payload operation commands, which contain a sequential list of commands with the date and time when each command will be issued.
The payload operation commands is uplinked to the satellite 1--2 times per week.

We can perform time of opportunity (ToO) observations for transient X-ray sources.
Changes in the attitude files for ToO observations follow the same procedure described above.
On the other hand, we do not need to change the payload operation commands because they are defined by only considering the radiation zone in the orbit.
The ToO response time is approximately one hour when we decide on a ToO observation just before a ground contact or 21~hr in the worst case.

In regular operations, downlink via the S-band is conducted three times per day. A daily payload data downlink of 60 MB is guaranteed, with any additional data being transmitted on a best-effort basis. 
The downlinked payload data and satellite telemetry data are transferred to Amazon Web Service (AWS) Cloud after the downlink and fetched from a server for payload operation installed at RIKEN. 
The binary data are decoded, processed using pipeline software, and visualized as time-series data with \texttt{Grafana}, an interactive web-based visualization application.
By reviewing time-series telemetry data, the NinjaSat science operations team monitors the satellite’s status and collaborates with the NanoAvionics operations team to resolve any issues that may arise.
The data output from the data analysis pipeline can be analyzed using standard X-ray data analysis tools such as \texttt{xspec}~\citep{Arnaud1996}.
Details of the data processing pipeline and data analysis are discussed in a separate paper.

By the end of 2024 November, we had observed 21 X-ray sources using NinjaSat:
Crab Nebula,
Sco X-1,
SRGA J144459.2$-$604207 \citep{2024ATel16495},
EXO 0748$-$676 \citep{2024ATel16678},
4U 1636$-$536,
GX 17+2,
Cyg X-1,
Cyg X-2,
MXB 1730$-$335,
Her X-1,
SMC X-1,
GX 301-2,
4U 0115+63,
1E 1841$-$045,
GX 339$-$4,
NGC 4151,
NGC 526,
T CrB,
Aql X-1,
GX 1+4,
MAXI J1752$-$457 \citep{2024ATel16903}, and 
NGC 4388.
The intensity of these sources ranges from faint (a few mCrab) to bright ($\sim$14~Crab).

\section{Concluding remarks}

NinjaSat is a 6U CubeSat designed for X-ray astronomy and represents the first observatory-type scientific CubeSat.
The NinjaSat project began in 2020, successfully launched on 2023 November 11, and achieved minimum success criterion on 2024 February 9. 
The X-ray detectors onboard NinjaSat remain in good condition, and the satellite has been operating smoothly with no major issues one year after the launch.
By the end of 2024 November, we had observed 21 X-ray sources using NinjaSat.

Although CubeSat missions carry a high risk of failure, effective risk management can enable cost-efficient satellite production and operations.
The X-ray detection sensitivity of NinjaSat is lower than that of large scientific satellites.
However, NinjaSat demonstrated that scientific results can be achieved by leveraging its ability to dedicate long-term observation to a single X-ray source \citep{takedaSRGAJ1444}.
We also demonstrated that, with a carefully designed time-tagging system using GPS, it is possible to achieve the timing accuracy of sub-millisecond, comparable to that of large satellites (sub-subsection~\ref{sec:gmc_timing}).
We will continue scientific operations aiming to achieve full and extra success.

The project name is inspired by "ninjas," the covert Japanese agents of the Warring States period.
Unlike the regular "samurai" army, ninjas operated in secrecy, were not bound by traditional rules, and employed advanced expertise.
Similarly, the NinjaSat project uses an agile CubeSat bus and flexible observation plan to pursue scientific discoveries in ways different from those of large, conventional satellites.

\begin{ack}
This project was primarily funded by RIKEN Cluster for Pioneering Research and partially supported by RIKEN Nishina Center. 
Additional support was provided by RIKEN Pioneering Project and JSPS KAKENHI (JP17K18776, JP18H04584, JP20H04743, and JP24K00673).
T.~E. was supported by the RIKEN ``Extreme Natural Phenomena'' Hakubi project.
T.~Takeda was supported by the JSPS Research Fellowships for Young Scientists.
N.~O. was supported by the RIKEN Junior Research Associate Program.
A.~A. was supported by the RIKEN Student Researcher Program.
We would like to express our gratitude to KEK-PF for their support with the X-ray calibration of the GMC, as well as to HIMAC and WERC for their assistance in radiation testing. 
Special thanks are extended to Y.~Maeda and Y.~Iwakura at ISAS/JAXA for their invaluable support during the thermal vacuum tests.
We also wish to thank K.~Yamaoka, K.~Nakazawa, and H.~Tajima at Nagoya University for their support in the thermal vacuum tests. 
Additionally, we acknowledge the support of TAC Inc. in the development of the NinjaSat electric boards.
Finally, we deeply appreciate the exceptional efforts of NanoAvionics and MBA in the production and operation of NinjaSat.
\end{ack}

\bibliography{ninjasat}
\bibliographystyle{apj}

\end{document}